# A Scenario for Strong Gravity in Particle Physics:
## An alternative mechanism for black holes to appear at accelerator experiments

D. G. Coyne[1]• and D. C. Cheng[2]

[1] *Institute for Particle Physics, University of California at Santa Cruz, Santa Cruz, CA 95064*
[2] *Almaden Research Center, 650 Harry Road, San Jose, CA 95120*

A different reason for the apparent weakness of the gravitational interaction is advanced, and its consequences for Hawking evaporation of a Schwarzschild black hole are investigated. Proceeding from some fundamental thermodynamic observations, a simple analytical formulation predicts that evaporating black holes will undergo a type of phase transition resulting in variously long-lived quantized objects of reasonable sizes, with normal thermodynamic properties and inherent duality characteristics. Speculations on the implications for particle physics are explored, and predictions for possible experimental confirmation of the scenario at LHC are made.

## I. MOTIVATION

In the quest for grand unification of particle physics and gravitational interactions, the vast difference in the scale of the forces, gravity in particular, has long been a puzzle. It has often been assumed that near the Planck scale, gravity would somehow assert itself and become comparable in strength to the other forces of nature, likely as a product of some grand unification picture. Most past attempts to reconcile high-energy particle physics (HEP) with general relativity (GR) have encountered severe technical difficulties. String theory, or other candidates for understanding Planck-level gravity, offer some hope for resolution of these difficulties, but characteristically give few if any explicit predictions for energy regions where experiments can sample particle physics or black hole astrophysics.

In recent years, string theory developments [1] have suggested that the "extra" dimensions of that theory, or special qualities of the "bulk" in those dimensions, are responsible for the weakness of observed gravity. While not a complete theory, such a scenario has interesting ramifications and even a prediction of sorts: if the extra dimensions have proper qualities, the Planck mass $M_P = \sqrt{\hbar c / G_N}$ effectively will be reduced to $\sqrt{\hbar c / G_b}$, $G_b$ being a bulk gravitational constant $\gg G_N$. Black hole production at the CERN Large Hadron Collider is then predicted [2], together with the inability to probe high-energy particle physics at still higher energies and smaller distances. Experiments searching for consequences of the extra dimensions have not yet shown evidence for their existence, but have set upper limits on their characteristic sizes [3]. Experiments at LHC aimed at detecting such black holes are being planned [3, 4], but suffer from not knowing exactly how a reduced-Planck-mass black hole would behave.

---

• deceased

In this paper we consider another approach to this problem of weak gravity and the merging of GR and HEP. Section 2 gives the methodology used in constructing the scenario and a preview of what we have found. Section 3 gives more details of the fundamental assumptions and conjectures made. Section 4 gives the analytical solutions of the model, the continuous states of which are then quantized in Section 5. The confrontation with experiments that soon may be performed is discussed in Section 6. Discussion and conclusions are left for Section 7.

## II.  APPROACH USED AND PREVIEW OF RESULTS

Past experience in physics suggests that when progress is arrested by "technical difficulties", it is often useful to use a well-established, more classical model for the physics involved, and then to modify it in the least intrusive manner consistent with experimental constraints, attempting to arrive at a less anomalous and more consistent picture of the underlying physics. For the near-Planck physics regime, we can think of no more appropriate a model than the now-classical model of black hole evaporation. The Bekenstein/Hawking/Page model [5], suitably modified [6, 7] for the Standard Model (SM), breaks down as anticipated near $M_P$ because it predicts horizonless singularities, infinite curvature and admits to an unknown back-reaction due to this curvature.

It is possible, of course, that both GR and quantum mechanics (QM) break down at this level. Our prejudice in this paper is that QM has proven reliable at the distance scales somewhat closer to those involved in the endpoints of BH evaporation, while the theory of gravity has proven its elegance on larger scales, and must be at least suspect at the scales needed for a complete description, including particle physics, at the end of the evaporation process.

In this paper, we consider whether there is any simple technique by which to modify the classical strength of gravity so as to pacify the anomalies of the evaporation process. Looking for hints from fundamental physics to keep this a less speculative search, we turn to basic thermodynamics, and constrain any modifications to the classical model to conform to experimental realities. We also consider a seemingly independent approach[1] that allows gravity to grow in strength only when the temperatures of evaporating black holes are trying to become infinite, and suggest a largely *ad hoc* picture of how a black hole might behave in this situation. These two lines of approach, used separately, both give strikingly similar results with a generic characteristic: a completely new dynamical solution for the evaporation ensues which regularizes the evaporation process. The reader may preview these results by examining Figs. 2-8, which show that the evaporation is free of physical infinities, possesses traditional thermodynamic properties after an apparent phase change, and likely conserves information. New insights into black hole thermodynamics can be made. Duality under the exchange $M/M_P \rightarrow M_P/M$ emerges naturally. The entire phase space of black hole masses can be quantized and testable predictions for black hole production at LHC are made: this scenario has substantially different predictions compared to the string theory models, and in particular predicts that there will be no end to the short distance probing of HEP almost down to the Planck length. It implies that while black hole production at LHC will likely occur, it could be in conventional 3+1 dimensions; furthermore, the process will appear as ordinary business-as-usual particle physics, with interesting new particles to be discovered.

The similarity of conclusions of these two new models we use may derive from one distinct commonality that they possess: they both predict that any states to be found at sub-Planckian masses will behave normally, i.e., will be essentially identical to elementary particles.[2] Perhaps the most reassuring conclusion that we find is that the dynamical solution in either model *forces* the sub-Planckian states to obey the Heisenberg uncertainty principle, and thus allows them to act as normal fundamental particles.

---

[1] This is the approach used in the two archival progress reports submitted earlier: <u>hep-th/0602183</u> and <u>hep-th/0609097</u>.

### III. THE FUNDAMENTAL ASSUMPTIONS OF THE SCENARIO

The key point made by the aforementioned string theory models is that gravity really isn't weak, but that the effects of strong (true) gravity are diluted by constraints felt by those who are confined to a lower-dimensional brane. In developing a new scenario, we use similar *initial* logic–that gravity is truly strong and fully comparable with other forces, but that we have not experimentally looked in those places where it resides. But instead of invoking extra dimensions and branes, we look elsewhere. An obvious place where experimentalists have not tested gravity, including GR, in any way, is directly at the horizons of black holes of sufficient temperature such that QG could be operative. We speculate that at this level, the spacetime structures of the horizons could be far more complex than those predicted by GR. They might well require more degrees of freedom to stipulate a particular state, and they might leak information; i.e., not be true horizons in the usual sense of the word. Most important, if gravity on or within these horizons is truly strong, yet we see no evidence of that on larger scales, then the complex horizons must be shielding in nature.

A shielding pseudo-horizon is a most unconventional speculation, but it has some historical antecedents [10]. First, a *gedanken geschichte*: imagine that the electrical force had been studied by (non-human) physicists who resided in a very cold universe where free electrons were not available. The Van der Waals forces between molecules would have characterized the electrical force, until eV scale accelerators were used to reveal the true strength and radial dependence of the interaction. While this might be a frivolous example, the true situation in nuclear physics was not so different. With only MeV accelerators, our universe was too cold for a deep examination of the strong force, and nuclear physicists had no idea, until reactions were done at higher temperature, that the force they had previously measured was actually only a remnant force between color singlets–the so-called nuclear Van der Waals force. The characterization of a strong, but asymptotically free interaction [11] came only later, in a locally hotter part of our universe. Note that the "anti-shielding" in this case came from previously unimagined gluons. The examples serve to reiterate our basic speculation. Our testing laboratories for the validity of GR in small black holes are far too cool, even frozen, compared to the appropriate temperatures.

We are, of course, aware of the usual objection to the idea of shielded gravity: that no antigravity source exists to do the shielding, unlike electric and color charges which have charge-conjugate states available for that function. We point out another obvious major objection: that attributing classical gravity only to the leakage from shielded small black holes seems impertinent, since gravity is such a universal phenomenon while small black holes are either rare or nonexistent [12]. These objections will be re-considered after the new scenario takes explicit form.

In order not to lose our way in a thicket of assumptions, we will cling to the basic formulations describing the interaction of black holes and particles–the modified Beckenstein-Hawking-Page formulation of Schwarzschild black hole evaporation in 3+1 dimensions. All the standard formulae will be taken to be valid at all temperatures and time scales. The effects of QG in the extreme regions will be taken to be lumped within the description of how Newton's *G* might be modified in those regions. We will select a form for $G(T)$ or $G(M)$ and then use it throughout, testing to see if the results are self-consistent and physically reasonable. All fundamental constants will be explicitly displayed in formulae to track the dependence on $G$ and to ease the interpretation of the new solutions.

---

[2] We have been greatly influenced by several authors who quite early arrived at this conclusion[8,9].

### A. Derivation of G from thermodynamic considerations

In attempting to find an anomaly-free model of black hole evaporation, good at arbitrarily small masses, we are equivalently searching for physics that is valid in the present regime of HEP. This suggests taking hints from existing knowledge that is either experimentally or theoretically on firm ground. One approach is to use the First Law of thermodynamics, together with the expected form of the entropy of the object being investigated (as a function of the energy of the object). For the simple Schwarzschild black hole, the First Law is [13,14]:

$$c^2 dM = dU = T_{Bk} dS_{Gibbs} = \frac{\hbar c^3}{8\pi GM} \frac{dS_{Gibbs}}{k_B} \equiv \frac{\hbar c^3 dS}{8\pi GM} \equiv \frac{\kappa dS}{GM}, \quad (3.1)$$

where $T_{Bk}$ is the (generalized) Beckenstein temperature [15] of the hole at a mass $M$ and any $G$. Note that $S \equiv S_{Gibbs}/k_B$ is dimensionless, expressible in nats or bits depending on whether the number of microstates involved is written as $e^S$ or $2^S$, respectively [16]. Then, letting $m \equiv M/M_P$ and $g \equiv G/G_N$, we rewrite **3.1** as

$$dS/dm = 8\pi gm \quad . \quad (3.2)$$

But there is also a second constraint on $dS/dm$, deducible from the expected form of $S$ for black holes at various mass scales. (It turns out that only two limiting black hole masses, very large and very small, need be considered to constrain $dS/dm$, for all practical considerations.)

*At large black hole masses*, far above $M_P$, it is well-established theoretically[3] that the entropy of the hole must be $S = $ Area/4, where the area $A$ of the black hole is measured in Planck area units $G_N \hbar/c^3$ [17]. Phrased in our definition above, the dimensionless entropy is

$$S(m \gg 1) = 4\pi m^2 + S_L \quad (3.3)$$

The additive arbitrary constant $S_L$ is usually chosen as zero, but is left arbitrary in this formulation.

*At low masses*, $M \ll M_P$, we look for an expression for entropy from QFT, but one which describes a single object which has a well-defined number of final states into which it can decay, as do heavier black holes. The pertinent form[4] seems to be S = log(number of microstates, or degeneracy) or S ∝ log(total energy of system) ∝ log(M), as suggested by Lykken [18]. This form requires a low-mass cutoff to avoid negative entropies, thus we take

$$S \propto \ln(M/M_L) = \ln(m/m_L), \quad (3.4)$$

---

[3] This is derivable from the First Law, if one assumes g=1 throughout this high mass region, which is experimentally confirmed by the observed constancy of $G=G_N$. I.e., this is an application of the First Law in a "frozen" universe.

[4] In a more general case of QFT, the log E dependence changes because the number of possible particles is not constant as above [19].

where $M_L = m_L M_P$ is the lowest-mass state of this neutral scalar particle-like black hole. The experimental constraint is that $M_L > \approx 100$ GeV [20].

*At intermediate masses*, $M < M_P$ but not $<< M_P$, $S$ is also expected to be proportional to log(number of microstates). In this region, string theory models often predict a "Hagedorn-like" exponential proliferation of string excitations with increasing energy, leading to $S \propto E$ [18]. For a new type of black hole state in this region we would then expect $S \propto M$. Our new scenario does not assume the validity of string theory or extra dimensions, but for generality we will add an arbitrarily weighted "stringy" term to the entropy having this dependence, turning on only at intermediate masses. When a solution is reached, we can examine the relevance of this term; it will be shown to be unimportant for feasible experimental checks of our scenario. It should also be noted that in different versions of string theory models, this Hagedorn region may or may not be present [19].

From the above discussion, we take as a constraint for our thermodynamically-constrained model:

$$S_{thermo} = A \ln(m/m_L) + B\sqrt{m^2 + C} + Dm^2 + S_L, \tag{3.5}$$

where $A, B$ and $D$ measure the relative importance of the terms in each energy region, while $\varepsilon \equiv \sqrt{C}$ gauges the onset (in energy) of the string-like linear term. The new constraint on $dS/dm$ is then:

$$dS_{thermo}/dm = A/m + Bm/\sqrt{m^2 + \varepsilon^2} + 2Dm. \tag{3.6}$$

Equating the two forms (3.2) and (3.6) for $dS/dm$, we then find a derived expression for $g(m)$:

$$g_{thermo}(m) = \frac{A}{8\pi m^2} + \frac{B}{8\pi\sqrt{m^2 + \varepsilon^2}} + \frac{D}{4\pi} \equiv \frac{a}{m^2} + \frac{b}{\sqrt{m^2 + \varepsilon^2}} + 2d. \tag{3.7}$$

Also constraining $d$ to match the theoretical value of $S$ at $M >> M_P$, we find $d = 1/2$, and

$$g_{thermo}(m) = 1 + \frac{a}{m^2} + \frac{b}{\sqrt{m^2 + \varepsilon^2}}. \tag{3.8}$$

### B. Derivation of G from anomaly suppression

An alternative method of generating the expected form of generalized $G$ is to simply pick a dependence guaranteed to avoid the anomaly of infinite temperatures and curvatures encountered in ordinary Hawking radiation formulae, once those formulae are applied outside the region of validity for which they were derived. Because $G_N$, to be replaced by variable $G$, appears in the denominator of the Bekenstein temperature, an *ad hoc* method of suppressing infinities is to add one! Namely, suppose we let $G$ approach infinity whenever the temperature of the black hole approaches the generalized Hawking temperature of a black hole at the Planck mass, here defined as:

$$T_P = \kappa / M_P G = \kappa / (G\sqrt{\hbar c/G}). \quad \kappa \text{ as defined in Eq. (3.1)} \tag{3.9}$$

Explicitly, to satisfy also the approach of $G$ to $G_N$ when temperatures are far below $T_P$, take

$$G = G_N T_P / (T_P - T). \tag{3.10}$$

It appears that we have introduced a pole in the definition of *G*, but since $T_P$ is a function of *G*, this implicit form is really a simple, if obscure, quadratic in √*G*. Examination of the solution for *G* reveals that 3.10 behaves locally as if *G* were approaching a pole, but where the pole has a variable value and moves **relative to** the current value of *T*. To avoid confusion[5], we henceforth add a label *vp* (for *variable pole*) to solutions based on this type of variable *G*. The expression is easily solved for *g(m)*:

$$g_{vp} = 1 + \frac{1}{2m^2}\left(1 + \sqrt{1+4m^2}\right) \tag{3.11}$$

Although the two expressions (3.8) and (3.11) came from completely different approaches, they have exactly the same asymptotic forms for very small *m* ($g \to 1/m^2$ if *a* =1) and for very large *m* ($g \to 1$). They have similar generic behavior in the intermediate region, with slight differences depending on the parameters *a* and *b* and even slighter differences depending on ε. The thermodynamic model can be considered to be a parametric generalization of the variable pole model. We will illustrate the modified Hawking evaporation using the thermodynamic model with *a* =1; *b* = 0.75, and then discuss the dependence of results on parameterization or the alternative model. One might ask if the variable pole form for $g_{vp}$ also reproduces an entropy similar to that from which $g_{thermo}$ was derived, and it does, with a somewhat different formula but generic similarity everywhere, and numerical similarity for the above parameters. From the thermodynamics of Eq. (3.2), substituting $g_{vp}$ and integrating, one finds:

$$S_{vp} = S_L + 4\pi\left\{2\ln\left(\frac{m}{m_L}\right) + (m^2 - m_L^2) + \sqrt{4m^2+1} - \sqrt{4m_L^2+1} + \ln\left[\left(\frac{1+\sqrt{4m_L^2+1}}{1+\sqrt{4m^2+1}}\right)\right]\right\} \tag{3.12}$$

The numerical similarity of Eq. (3.5) and Eq. (3.12) (see Fig. 9) suggests again that we have found two routes to the same physics. For both models, all results are calculable analytically; no numerical integrations are necessary.

## IV.  ANALYTICAL SOLUTIONS

At his point we have a thermodynamically-inspired modification to $G_N$ for use over the entire mass range of black holes, and a similar modification from the *ad hoc* arguments. Many authors have speculated on various changes in *G* to accommodate astrophysical/cosmological discoveries; we do not. Either of the two g(m) found above gives results identical to GR on all scales appropriate to those discoveries. The differences in our approach manifest themselves only at Planckian and sub-Planckian black hole masses. However, the general logic of modifying Newton's constant to approximate general features of a deeper underlying theory has been justified by a number of authors [22]. It now remains to test our form of *g(m)* by using it to modify the standard Beckenstein-Hawking-Page formulae.

We use a form of the evaporation equations suggested by Carr, MacGibbon, Halzen and Zas [6,7] which includes the most modern version of the Standard Model (SM) of particle interactions, while keeping the essential form of the classical equations invariant. Hawking's result [23], expanded upon by Page [24], was that an evaporating Schwarzschild black hole behaves like a slightly modified black body radiator, and obeys a modified Stefan-Boltzmann law: Power = $\sigma_B T^4 A$, with temperature $T$ as in Eq. (3.1). Rewritten explicitly for a conventional black hole:

$$\frac{d(Mc^2)}{dt} = -\sigma_B \left(\frac{\hbar c^3}{8\pi k_B G_N M}\right)^4 4\pi (2G_N M/c^2)^2 \quad (4.1)$$

---

[5] After writing this paper, we found that at least since 2005, "anomaly suppression" has been used to designate the very beautiful derivation of the Hawking radiation formulae by imposing side conditions on GR to avoid conflicts with QM. [21]. Though much more sophisticated, those techniques are not so different in spirit than what we attempt here, although with completely different results. Because of this, we use a different name for the technique of **3.9**.

However, the size of the radiator may be small compared to the emitted wavelengths, leading to spin-suppression factors in $\sigma_B$. In addition, the evaporating black hole can produce many more degrees of freedom than the two polarization states of a photon referenced in the conventional Stefan-Boltzmann constant. With these corrections and the lumping of constants, the equation becomes

$$dM/dt = -\alpha(M, spin)/M^2 \equiv -\beta(M, spin)/G_N^2 M^2, \quad (4.2)$$

where $\alpha$ (or $\beta$) include all the correction factors for particles which can be produced consistent with the temperature scale $T$ of the hole. $\alpha(M)$ is shown in Fig. 1, and in the SM becomes essentially constant at $\alpha_o \approx 7.8 \cdot 10^{26} g^3/s$ by the time a hole has evaporated down to about 1000 metric tons of mass and is producing a putative Higgs boson, along with all less massive elementary particles, in its evaporation radiation. Note $\beta_0 = \alpha_0 G_N^2$.

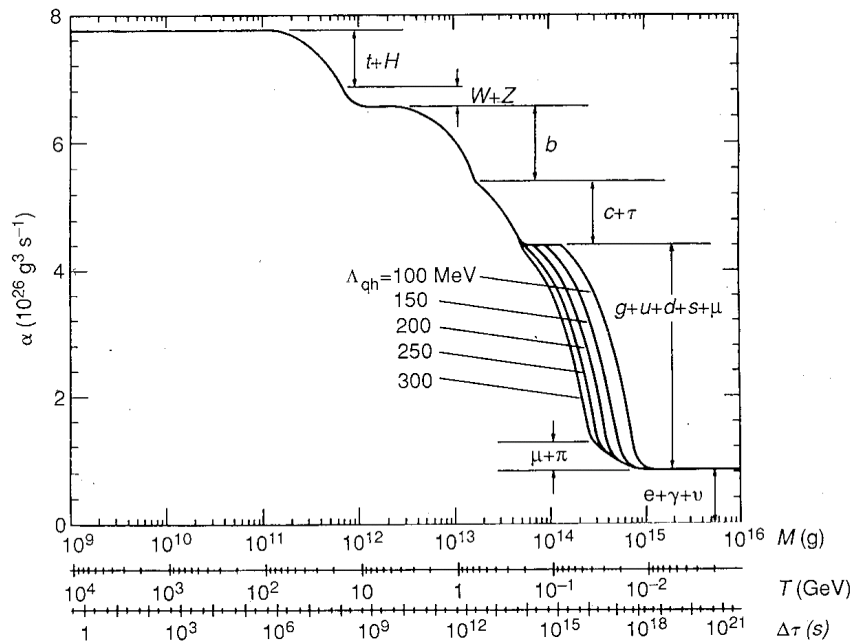

FIG. 1. [Halzen, et al, 7]: *The parameter $\alpha$ counting the degrees of freedom of the Hawking mass radiation as a function of the black-hole temperature and lifetime. $\Lambda_{qh}$ is the quark-hadron deconfinement scale. The contribution from each particle species is indicated.*

Moreover, the standard Hawking solution: $\dfrac{\alpha_0 t}{M_P^3} = -\int_{m_0}^{m} m^2 dm$ (in our *m*-notation) (4.3)

shows that if future groups of higher mass particles are encountered anywhere in HEP, and the hole with increasing temperature becomes capable of emitting them, only the time scale $\alpha_0 t$ is changed, and not the essential features and problems of Hawking evaporation. In the "desert" hypothesis, where few new particles appear up to the Planck mass, $\alpha_0$ is expected to stay almost unchanged. It would double with the onset of SUSY, and decrease if there were a substructure to quarks involving fewer degrees of freedom [25].

With our generic scenario, which we label SSGS for brevity (shielded strong gravity scenario), the revised solution to the evaporation equations is simply:

$$\frac{\alpha_0 t}{M_P^3} = -\int_{m_0}^{m} m^2 g^2 dm, \quad \text{with } g(m) \text{ as in Eq. (3.8), or (3.11).} \tag{4.4}$$

This integral is analytically solvable in both cases and the time solution using **3.8** is:

$$t - t_0 = \frac{M_P^3}{\alpha_0} \left\{ \frac{a^2}{m} - b^2 \left[ m - \varepsilon \arctan\left(\frac{m}{\varepsilon}\right) \right] - \frac{m^3}{3} - b(2a - \varepsilon^2)\ln(m + \sqrt{m^2 + \varepsilon^2}) - bm\sqrt{m^2 + \varepsilon^2} - 2am \right\}$$
$$- \left\{ \text{same expression evaluated at initial mass } m = m_0 \right\} \tag{4.5}$$

Above, $M_P^3/\alpha_0 \approx 1.322 \cdot 10^{-41}$ seconds. Then, from the Eq. (3.8) above, this time solution for m allows us to compute $g(t)$, as well as the horizon radius:

$$r(t) = 2mg \quad \text{(in Planck lengths);} \tag{4.6}$$

and the temperature: $\quad k_B T(t) = T_{PH}/mg$. $\tag{4.7}$

$T_{PH}$ is defined as the temperature predicted by the conventional Hawking treatment at a black hole mass of (fixed) $M_P$, or $T_{PH} \approx 4.85 \cdot 10^{17}$ GeV.

### A. Details of the solution for the thermodynamic model

Although Eq. (4.5) is not transparent, a few general features of the solution are evident by inspection of Eqs. (4.4) and (3.8). Since $\alpha$, $m$ and $g$ are all positive definite, $m(t)$ can only decrease with time. Because $g \approx 1$ for large $m$, a typical solution starting at large initial mass begins by mimicking the standard Hawking behavior, as if the object had a finite life. But for very small $m$, one sees that $g \to 1/m^2$ and thus, using Eq. (4.4), $m \to \propto 1/t$: the black hole will never evaporate and can reach arbitrarily small masses. $g(t)$ grows quadratically with time, never reaching a pole: it is dynamically suppressing its own infinities. Temperature $T \propto 1/t$ as well in this limit: the black hole cools off! (A display of these and other asymptotic behaviors that demonstrate the inherent duality of the model under the exchange $M \leftrightarrow 1/M$ can be found in Appendix A). Clearly, something interesting has happened in the intermediate region where Hawking behavior has been transcended.

From the solution (4.5), the following Figs. 2-8 illustrate these initial conclusions and introduce new surprises. These figures do not display the vast majority of the time scale (for $m_0$ starting $\gg 1$), because the solution in that time region is indistinguishable from the Hawking case. Instead, a typical figure begins at $m_0 = 3$, where a sudden change from the traditional behavior soon ensues. Note that the time intervals of pertinence are very small, of $\mathcal{O}(10^{-40}\text{s})$, but are still $\gg \tau_P$

≈ $10^{-44}$ s by four orders of magnitude. Fig. 2 below compares a particular case of the thermodynamic model with the traditional Hawking solution, showing that the Hawking pole disappears, the mass evaporates smoothly through the Planck mass, and the temperature indeed peaks as the hole reaches the Planck mass (at about $30 \cdot 10^{-41}$s) and cools off thereafter. Note that this does not negate the astrophysical black hole explosions proposed by Page and Hawking—they simply are "all over" by the time the remnant approaches the Planck mass. It does imply, however, that some tiny amount of mass does not partake in the explosion.

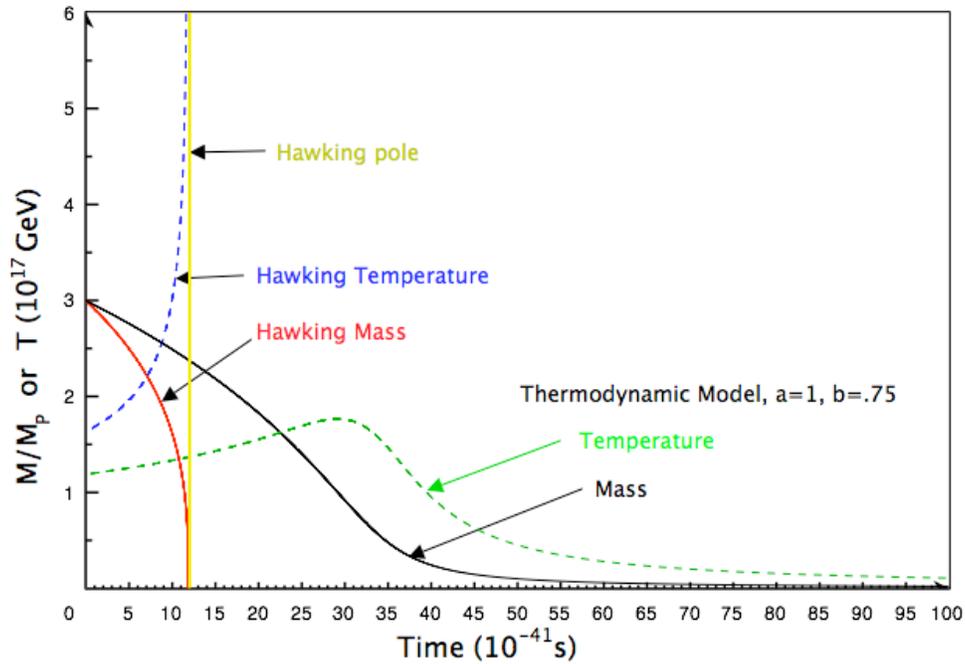

*FIG. 2. Comparison of the thermodynamic model with the Hawking model near $M_P$*

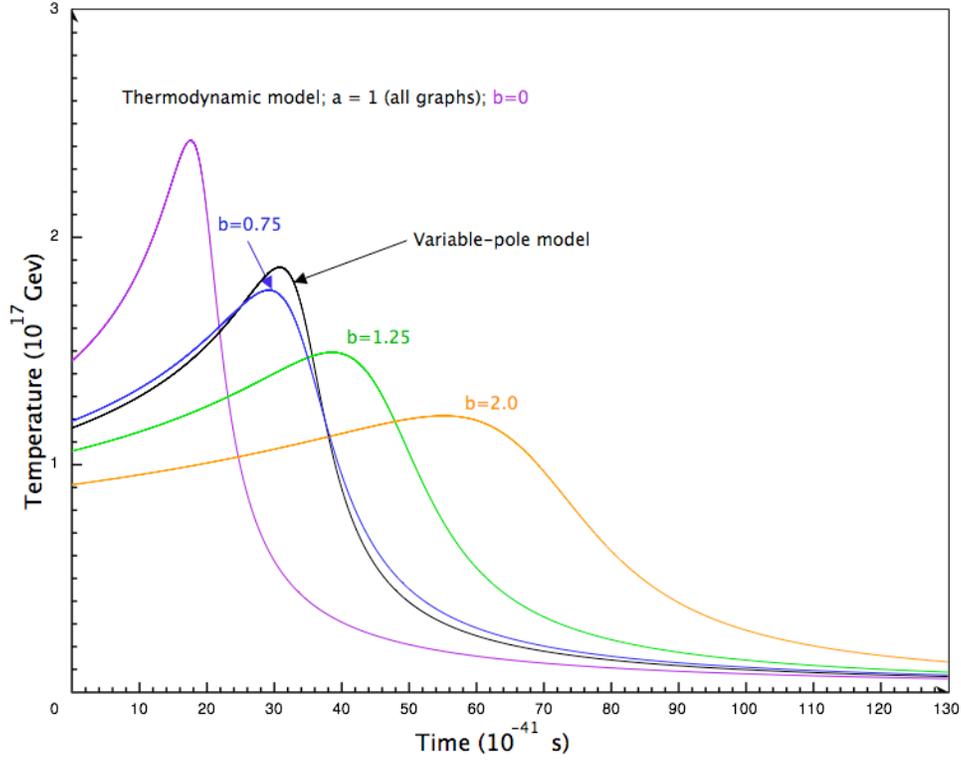

*FIG. 3. The time dependence of the temperature in the two SSGS models*

In Fig. 3 above, the temperature dependence on time is displayed for models based on either $g_{vp}$ or $g_{thermo}$, with a range of the $b$ parameter for the latter. ($a$ is fixed at 1, and the $\varepsilon$ parameter is fixed at $\varepsilon M_P$=1 TeV, but its variation has little or no effect on the curves). The variation of $b$ does not cause any change in *generic* behavior, but has an effect on how sharp the transition is near the Planck mass. In every case (and for the other physical variables of the black hole evaporation), the solutions for the two new models quickly converge to identical asymptotic forms, both for high mass and for tiny mass regimes. For this reason, we henceforth choose to display other dynamical variables and predictions using only the model based on $g_{vp}$, which has no parameters and typifies the generic behavior.

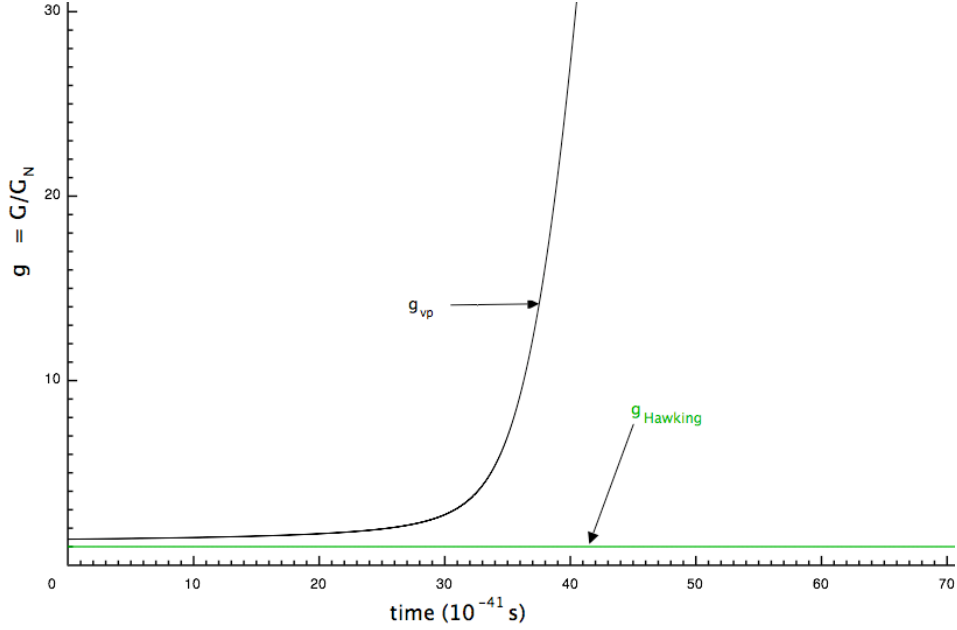

*FIG. 4. Comparison of the time behavior of G in the variable pole and Hawking models.*

Consider now the behavior of $G$ (Fig. 4). Hovering at a few times $G_N$ up to the critical transition, $G$ starts to climb. At this transition point it is roughly $5G_N$. From Eq. (3.11), $g_{thermo}$ has horrific positive exponents by the time m reaches the locale of familiar elementary particles, but we will see that they are just what is necessary to make gravity comparable to other interactions. This quantitative treatment has shown that a dynamic model of gravity gives it the possibility of different strengths at different mass scales. For instance, a "black hole electron" internal state (or intra-horizon state) would be characterized by a much stronger coupling $G_e$ instead of $G_N$, ignoring for the moment that the electron is not a neutral scalar. This modifies the usual statement of relative strengths of interactions, typically stated in the comparison of Coulomb and gravitational forces:

$$F_{gravity}/F_{Coulomb} \approx G_N M_e^2 / (\alpha \hbar c) = 2.4 \cdot 10^{-43}$$

(here $\alpha$ is now the usual fine-structure constant). The variable $G$-pole model asymptotically has $G_e / G_N \rightarrow (M_P / M_e)^2$, and consequently, if particles could interact with their internal scale of gravity:

$$F_{gravity}/F_{Coulomb} \approx G_e M_e^2 / (\alpha \hbar c) = G_N M_P^2 / (\alpha \hbar c) \approx 137.$$

Clearly, $G$ has climbed to just what it needs to be to considered a strong force in HEP. We will return in Sec. 7 to a discussion of what this could mean for particle physics.

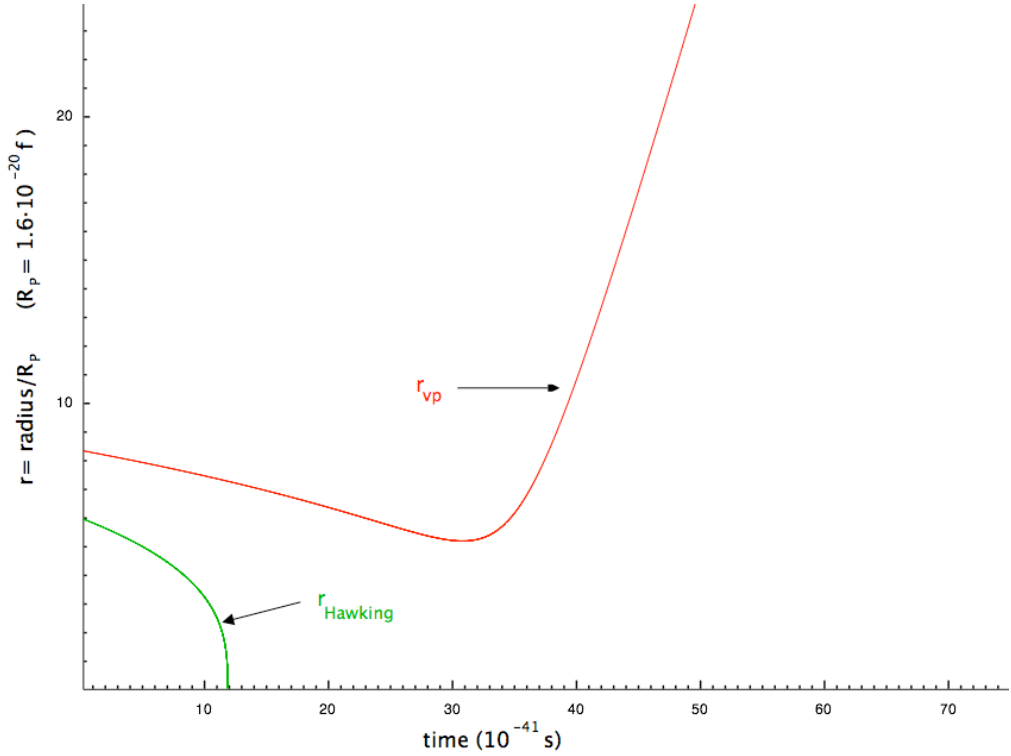

*FIG. 5. Comparison of horizon radius for Hawking and variable-pole models*

An unanticipated surprise in the time solutions is the behavior of the black hole horizon (Fig. 5), which of course in the old Hawking solution would quickly go to zero radius, introducing conceptual problems: an object being incredibly smaller than its Compton wavelength; a naked singularity; possibly lost information. Here, the solution to this model halts the shrinkage of the horizon at a radius about five times the Hawking radius at the Planck mass, and then proceeds to re-inflate the horizon! Since $r$ is proportional to $gm$, its asymptotic form in this time region is $\propto t^2 \cdot (1/t) \propto t$. What is the significance of this linear inflation of the "new horizon"? Fig. 6 shows that at the transition time the quantity $gm^2$ becomes essentially constant at the value 1. Since $gm^2 = GM^2/\hbar c$, this is exactly what is needed to keep the horizon radius $2GM/c^2$ greater than the Compton wavelength $\lambdabar = \hbar/Mc$ of the remaining black hole:

$$2GM/c^2 > \hbar/Mc \Rightarrow GM^2 > \hbar c/2$$

Thus, after the transition, the physical size of the object never violates a reasonable limit, quite unlike the constant $G_N$ Hawking solution. This guarantees that the remnant object, whatever it may be, will satisfy the uncertainty principle just as does any fundamental particle.

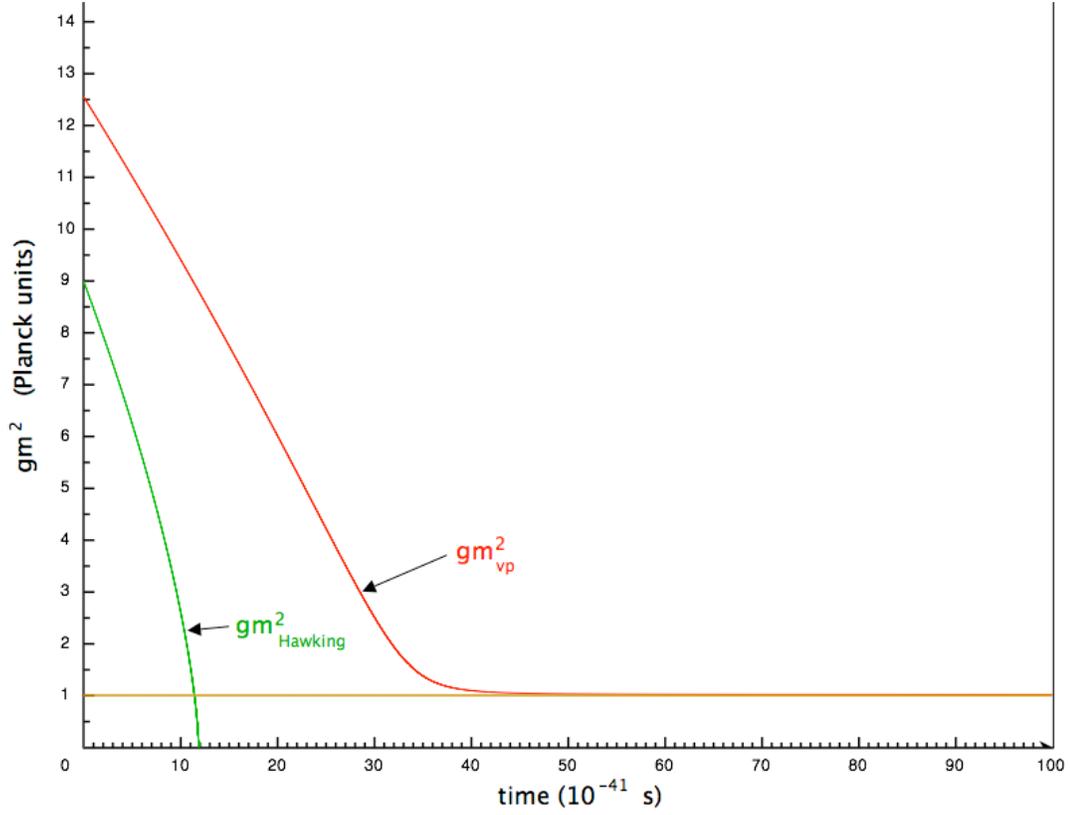

*FIG. 6. Comparison of gm² for Hawking and variable-pole models*

There is even more significance to the inflation of the horizon. Fig. 7 shows the time evolution of the embedding diagram of this system relative to that of a traditional Hawking solution. (Hawking solution at left, SSGS solution at right; the radial direction is in the x-y horizontal plane, with the diagram terminating at the horizon. The embedding "funnel" is also cut off at an arbitrary positive constant of z and r to better illustrate its shape). At a mass $10\,M_P$, (Fig. 7a) the solutions are indistinguishable. The horizon of the SSGS solution shrinks more slowly than does that of the Hawking solution, as $M \rightarrow M_P$, and its corresponding potential well at a given radius is deeper[6].

---

[6] The embedded surface is *not* the effective two-dimensional potential, but is intimately related to it through the appearance of the lapse function both in the metric and the potential [26]. Plots of the potential look very similar to these plots, so would be redundant here.

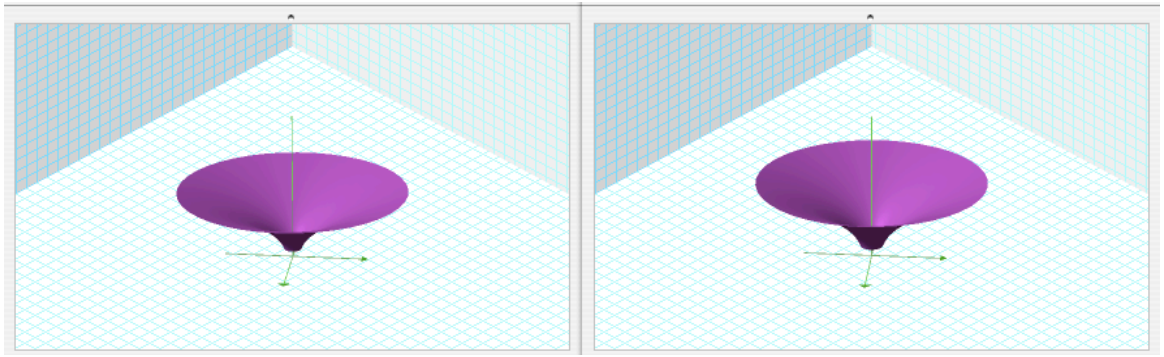

FIG. 7a.  $M = 10M_P$

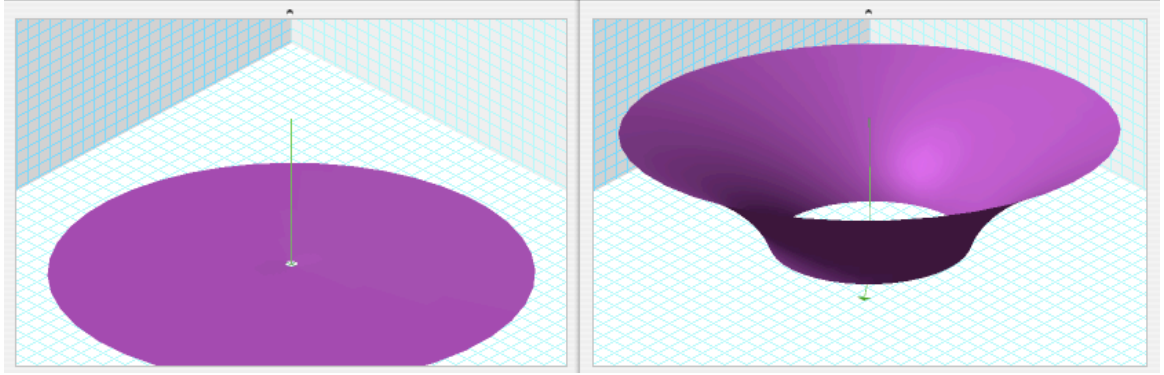

FIG. 7b.  $M = 0.025M_P$

*FIG. 7.  Time evolution of the embedding diagrams for Hawking and SSGS models (see text for details)*

By the time the mass has shrunk to .025 $M_P$, (Fig. 7b), the SSGS solution has "bounced" and the duality of that solution is manifest, while in the Hawking solution the potential well and horizon have almost disappeared and the singularity is well on the way to becoming "naked".  The SSGS model inflates the horizon and keeps the singularity "clothed", within a deep potential well.  Depending on the nature of the horizon and its interior, quantized stationary states are at least plausible, as will become more evident as the model is investigated.  It is these two key features that will help make the model consistent with the idea that a shielded large gravitational force could lead to particle-like final states.

It is important to note that while variation of parameters in some SSGS solutions can cause significant changes in the time-dependent solutions, usually in the form of time displacements at the Planck scale, the fundamental physics quantities all seem to have the same generic behavior throughout.  Even more important, the regions of mass where there is any hope for experimental verification are either above ~$10^{16}$ $M_P$ or below ~$10^{-16}$ $M_P$, and in these regions all solutions of the two new models both converge to the same answers.

### B. Thermodynamics of the variable pole model:  specific heats and entropy

Additional insights to this solution can be gained by examining the behavior of the evaporation as a non-equilibrium thermodynamic system. Using Eq. (3.1), we first calculate the specific heat $c_{BH} = dU/(UdT)$.  The thermodynamic properties revealed will be grossly different than for a traditional Hawking evaporation, as has already been demonstrated for the temperature. The specific heat of a classical black hole is always negative [27], as shown in Fig. 8 (open arrowheads indicate the direction of time evolution in the evaporation process). However, this new type of black hole evaporation has a surprise: the specific heat has a pole in $U$ (and in $T$) at the "transition" temperature, and, after the infinite discontinuity, a second branch in which its specific heat is normal and positive (and where the black hole can be in equilibrium with a thermal bath). It is tempting to interpret this discontinuity as a phase transition, but from what, into what? There have been conjectures that black hole horizons really have substance, as found in models in which they are emergent phase transitions of the vacuum [28], or simply shells of emitted particles [10]. Here, all that can be deduced is that there is a critical point where the SSGS model ceases to follow the unusual thermodynamics of the classical black hole, and instead produces a state that looks more and more like an object obeying traditional thermal physics. There is no infinity as m goes to zero, because the model, with its generic logarithmic entropy, has a lowest mass cutoff at $m_L$. The infinities at the pole are not essential, in that they simply correspond to the maximum in the temperature as the mass continues to decrease—a consequence of increasing $g$. In Sec. 5, we derive quantized states of the system, which show that the hole actually spends a short but finite time (of order $10^4$ Planck times) at this transition temperature, suggesting that there is something like a latent heat involved.

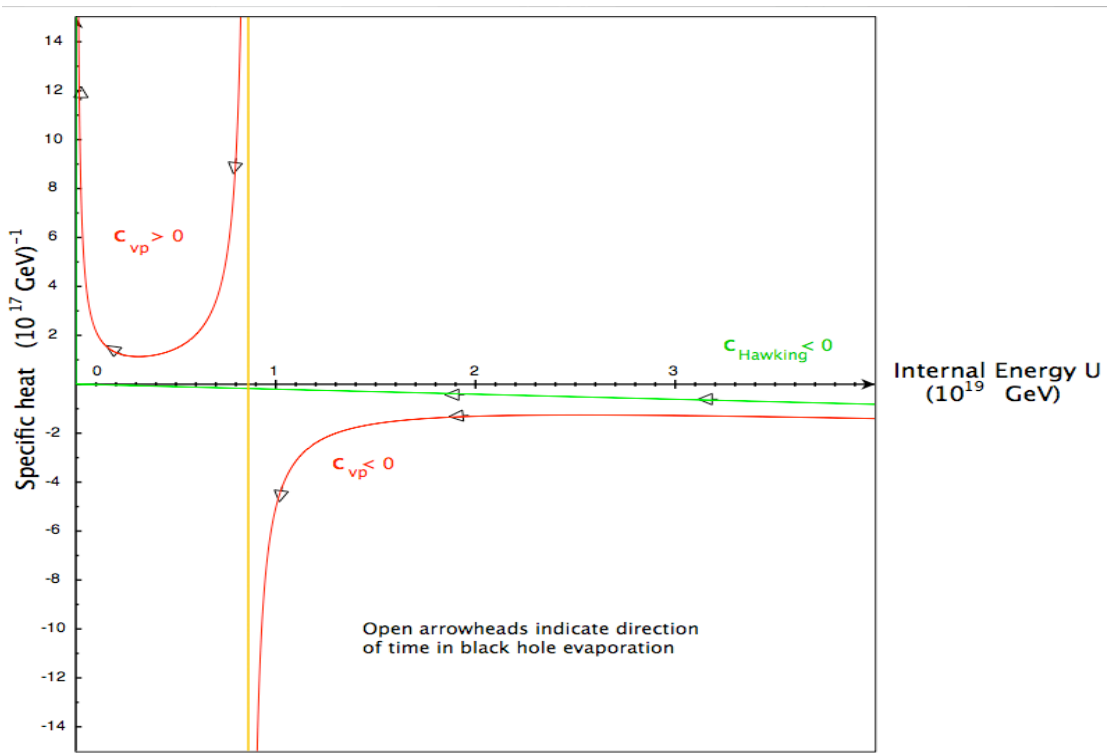

FIG. 8. *Comparison of specific heats for the Hawking and variable pole models*

### C. Entropy

In any discussion of *classical* black hole thermodynamics, entropy is considered to be one of the most important properties, given its close connection to information flow and the quantum mechanical concept of pure states [29]. From thermodynamic analogies, the classical entropy was expected to be $S = c^3 A/(4G\hbar)$ [$A$ = horizon area], and thus equal to $4\pi M^2 G/(\hbar c)$, taken as a measure of the concealed information within or on the horizon. The classical entropy of black holes is without experimental confirmation, but is so well-established theoretically that the agreement (in special cases) of the string theory calculation [30] with the standard formula has been taken to be a major triumph for that theory. Numerical evaluation of the formula reveals that entropy can be huge for astrophysical black holes: a solar-mass black hole has $S = 1.05 \cdot 10^{77}$, a factor of $10^{19}$ greater than the entropy of the Sun itself [31]. By contrast, the same formula, applied much lower than its expected region of validity, gives the entropy of a Hawking black hole of Planck mass to be $4\pi$, tending to zero like $(M/M_P)$.

In the SSGS models, we have already guaranteed that in the thermodynamically-inspired derivation of $g_{thermo}(m)$, the entropy (3.5) will reproduce the general expectations for "established" physics, at least within proper choices of the parameters $a$, $b$ and $m_L$. This includes the high mass classical limit just discussed. Note however that the anomaly-canceling (variable pole) version of SSGS associated with $g_{vp}$ also accomplishes this goal automatically, with no thermodynamic inputs and with only $m_L$ as free parameter in Eq. (3.12). In Fig. 9 we display a variety of solutions; all curves are shown with $m_L$ arbitrarily set to the lowest value, 100 GeV, consistent with present experiments, and with $S_L$ arbitrarily but consistently set to 11.09 nats (16 bits). First we compare the entropy from the $g_{thermo}(m)$ model for $a = 1$, $b = .75$, $\varepsilon = 10^{-16}$ to the model derived from $g_{vp}$; they are identical within the width of the plotted curve. This is also true for b=0 and a range of b of order unity. To illustrate the further lack of dependence on the b parameter, we include a curve with $a = 1$, $b = 1000$ and $\varepsilon = 10^{-16}$, showing that it is difficult to induce a "stringy" term without a huge "b" value. Even then, it affects the entropy only near the Planck mass, and is not important for practical experimental tests. Finally, for both these curves, we vary the parameter $\varepsilon$ (which determines an energy where any "stringy" term could start to contribute) by a factor of $10^{10}$, and find this makes no visible difference at all. The derived formulae for entropy are thus exceedingly stable in the regions of experimental interest, and this helps stabilize our later predictions for experiments at the LHC. In particular, the SSGS models show an entropy with an extensive logarithmic plateau, in striking contrast to the classical Bekenstein-Hawking entropy. This will be of enormous consequence in Sec. 5.

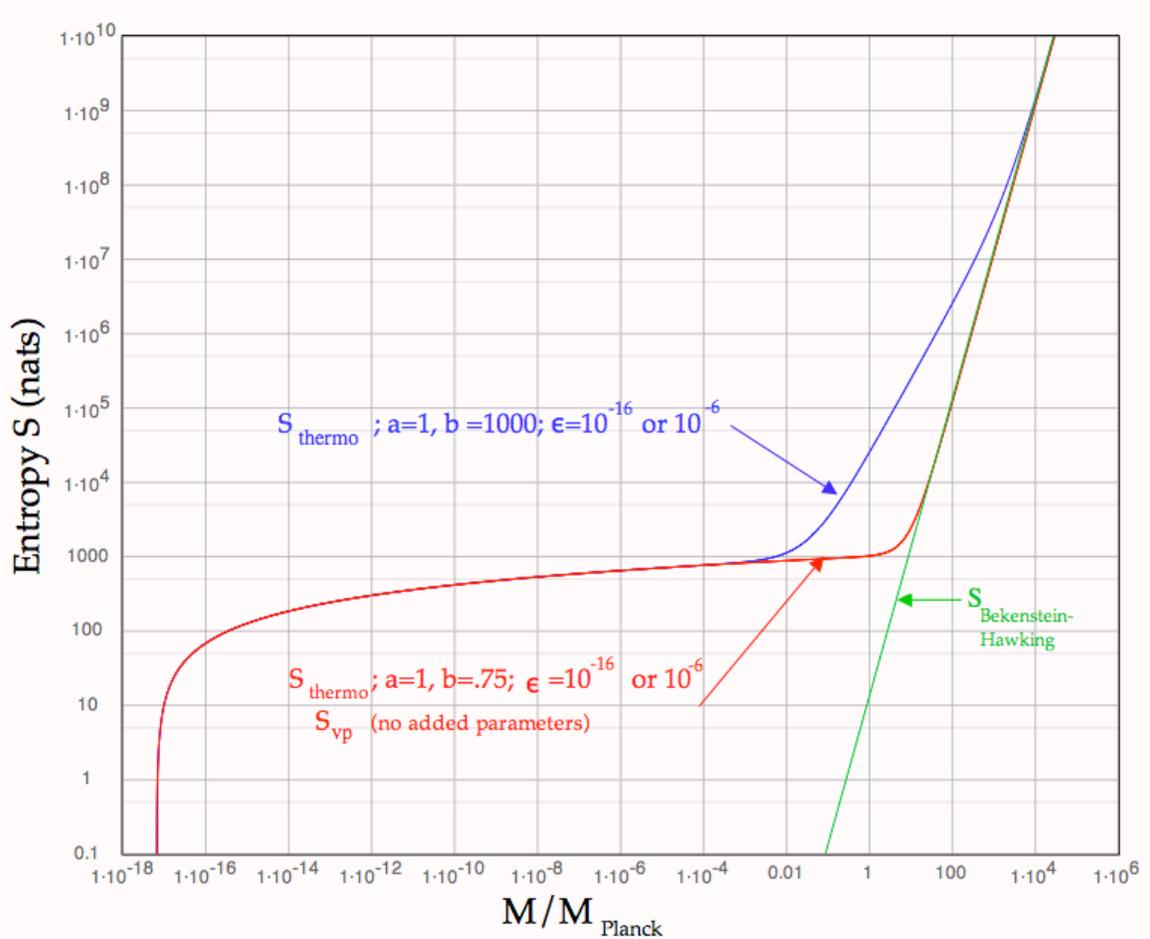

FIG. 9. *At least seven variations of the entropy of the SSGS models, plus that of the Bekenstein-Hawking model (see text for full explanation).*

## D. A conjectured new version of the Second Law of Thermodynamics

Thus, SSGS predicts that sub-Planckian objects have appreciable intrinsic entropy, and much more so than extensions of the classical model. Even at $M = M_p$, it exceeds the (dubious) classical value by two orders of magnitude (Fig. 9). This might be thought to be unimportant in the super-Planckian physics of black holes, where the additive constant (about 1000) is completely trivial. But the view espoused here—that the sub-Planckian objects are particles—leads to the conclusion that large black holes with entropy decay into many sub-Planckian ones, which also have intrinsic entropy. From a calculation in appendix B, we find that the logarithmic behavior discovered for entropy in the sub-Planckian region, coupled with the usual behavior above the Planck mass, effectively *guarantees* the validity of the second law of thermodynamics, *just* for the intrinsic entropy of black holes. Fig. 18 in that appendix shows that a neutral scalar black hole having any mass, super- or sub-Planckian, will rapidly transfer its intrinsic entropy to a very large number of sub-Planckian substates, with a net increase in the total sum of intrinsic entropy. If the intrinsic entropies involved are entanglement entropies, then this demonstration must bear on the infamous "information problem", even though many believe that has been partially or completely solved by other considerations [32]. The argument is not completely tight, and has a few caveats, so we regard its validity as only an interesting conjecture, following from a generalization of the concept of particles as black holes.

In discussions of black hole entropy and quantum black holes, attention is often centered on the entropy per unit area as a critical parameter [33], related to observance of the Bekenstein Bound [34]. The bound is derived in a weak-gravity limit, so may not be appropriate in the SSGS context [35]. For the classic GR black hole, $S/A = c^3/(4G_N \hbar)$ is constant at 1/4 nats/Planck area, and saturates the Bound. QFT predicts infinite entropy density, considered a disaster for that theory [29]. For regions between the two asymptotic regions and near the Planck mass, the SSGS model shows a small bump in entropy density, which is of unknown significance; it briefly (and modestly) over-saturates the Bekenstein Bound. For $M > 10 M_P$, SSGS gets the GR result, but for $M << M_P$ it is:

$$\frac{S}{A} = \frac{8\pi \ln \frac{M}{M_L}}{4\pi \left(\frac{2MG}{c^2}\right)^2} \to \frac{m^2}{2} \ln\left(\frac{m}{m_L}\right) \text{ in Planck area units}, \quad (4.6)$$

which is far less than the Bound for m < 0.1. As $M \to 0$, $G \to G_N M_P^2/M^2$ and the horizon inflates, it appears that black holes as elementary particles have very sparse information content per horizon area (becoming arbitrarily close to zero), but appreciable absolute information content. For example, a scalar black hole stabilized at $M = 1$ TeV would have $S \approx 87$ bits, allowing (but not requiring) about $10^{26}$ possible internal configurations of the microstates of its (presumed) shielding horizon, all within a radius of about $4 \cdot 10^{-3}$ f. Because of the logarithmic dependence, all other particles, from electron-like masses almost up to the Planck scale, would have similarly large numbers of microstates; this would seem to provide adequate phase space for scalar fundamental particles with widely differing properties and enormously complex horizons. (At the Planck mass, the 1440 bits provide $10^{434}$ possible configurations).

## V. QUANTIZATION OF STATES; LIFETIMES

Thus far, in this paper, the SSGS has provided considerable detail on the evaporation characteristics of Schwarzschild black holes. The natural question arises: how can this scenario be tested? If, as SSGS predicts, LHC physics will be largely business-as-usual (see Sec. 6), that would hardly elicit belief in this new scenario. To make matters worse, astrophysical black holes are not likely to be found close enough for experiments to follow the evolution of their final throes of evaporation. The total evaporation of an $M_P$ black hole releases only $10^{16}$ ergs, while the search for exploding primordial black holes, which release $10^{30}$ ergs in the last second of life, has remained fruitless [36]. This suggests not many such objects even exist. Traditionally, hypothesized objects are discovered by production of their quantum states and the transition radiation among them; are these continuously evaporating black holes likely to possess such quantized states?

### A. Quantization

It may come as a surprise to some that the quantization procedure for black holes has been a cottage industry for about thirty years, beginning with Bekenstein's use of "adiabatic invariants" to justify the quantization of black hole horizon area [37]. A wide range of theoretical speculation has expanded on such ideas [38,39,40], and there is even a lively discussion in current literature arguing the fine points of exactly what the fundamental quantum of black hole area should be [41,42,43,44]. A commonly occurring value [45] is

$$A_0 = 4\ln 2(\hbar G_N/c^3), \text{ or } 4\ln 2 \text{ in Planck area units.} \tag{5.1}$$

For classical Hawking evaporation, the actual mass states are easily derived from this assumption. Since in general

$$A_{bh} = 4\pi(2GM/c^2)^2 \text{ or } 16\pi(gm)^2 \text{ in Planck units,} \tag{5.2}$$

then one derives $\Delta A_{bh} \propto M\Delta M$, and with $\Delta A_{bh}$ quantized in units of $A_0$, $\Delta M \propto 1/M$. The classical ($g = 1$) mass states themselves, all for masses large compared to $M_P$, are:

$$M_n = M_P\sqrt{1 + \frac{(n-1)A_0}{16\pi}}, \quad \text{for large } n \text{ and with } A_0 \text{ in Planck area units.} \tag{5.3}$$

Predictions have been made [39] that TeV gamma rays will be produced from a $7\cdot 10^9$ g black hole transitioning between such states; such a hole would have about 100 s to live and would be rather dim. Badly smeared spectral lines would be even more difficult to detect than the aforementioned explosion with most of its energy release in the last second.

For the SSGS model, however, a new feature becomes evident and permits practical detection of quantized states. Unlike the models of the references, the SSGS follows the black hole evaporation not only down to $M_P$, but though that mass and as far down as the lowest state suggested by the entropy model, $M_L >$ about 100 GeV. Quantized black holes might be detectable by standard HEP techniques. To quantify this expectation, we build our own model of quantization, which follows.

At the very outset, we have a dilemma: As described, both in classical models and in SSGS (for large masses), $S = A/4$ (units of $S$ are nats and those of $A$ are Planck areas). Quantization of $A$ (in that mass region) with $A_0 = 4\ln 2$ is exactly equivalent to that of $S$ with $S_0 = \ln 2$. The underlying physical picture, however, can be quite different. While area quantization is a geometrical sizing of the horizon, entropy quantization attributes a different information content to each step of horizon size, representing the unseen microstates of that horizon—reflecting the degeneracy of the state or how many ways in which it could have been "constructed". Both long ago and more recently [46], J. A. Wheeler argued persuasively that *information* is the fundamental reality of the world. The SSGS model is not so extreme, but does attribute information in the complicated shielding horizons as a source of many degrees of freedom; entropy-quantization is the more natural scheme to use. But not to leave any stone unturned, we check both schemes for their internal consistency.

All this may seem like a non-issue, since in the super-Planckian region the results must be identical for the two quantization schemes. In the sub-Planckian region, however, the SSGS predicts hugely different results for area or entropy quantization—different not only from the classical picture, but from one another. Fig. 10 shows this in a very simple way. A grid of quantized steps of either $S$ or $A/4$ can be superposed on the vertical axis; we have used an arbitrary one with steps of 90 nats for pictorial clarity. Looking to see where these horizontal lines intersect the $S$ or $A/4$ curves then determines the corresponding quantized mass; the huge differences between the two quantization schemes are immediately clear. For area quantization, there is a natural lowest value of area ($A_{min}$), and there are pairs of states roughly symmetrical (on a log plot) around this point. In our notation, quantum numbers are positive above this point, and mass increases as quantum number increases from $n=1$. Below this point, quantum numbers are negative, and mass decreases as quantum number increases negatively from $n=-1$.

Even with the large quantum jumps of Fig. 10, the area-quantized states become very dense on the mass scale as larger (absolute) quantum numbers are reached.

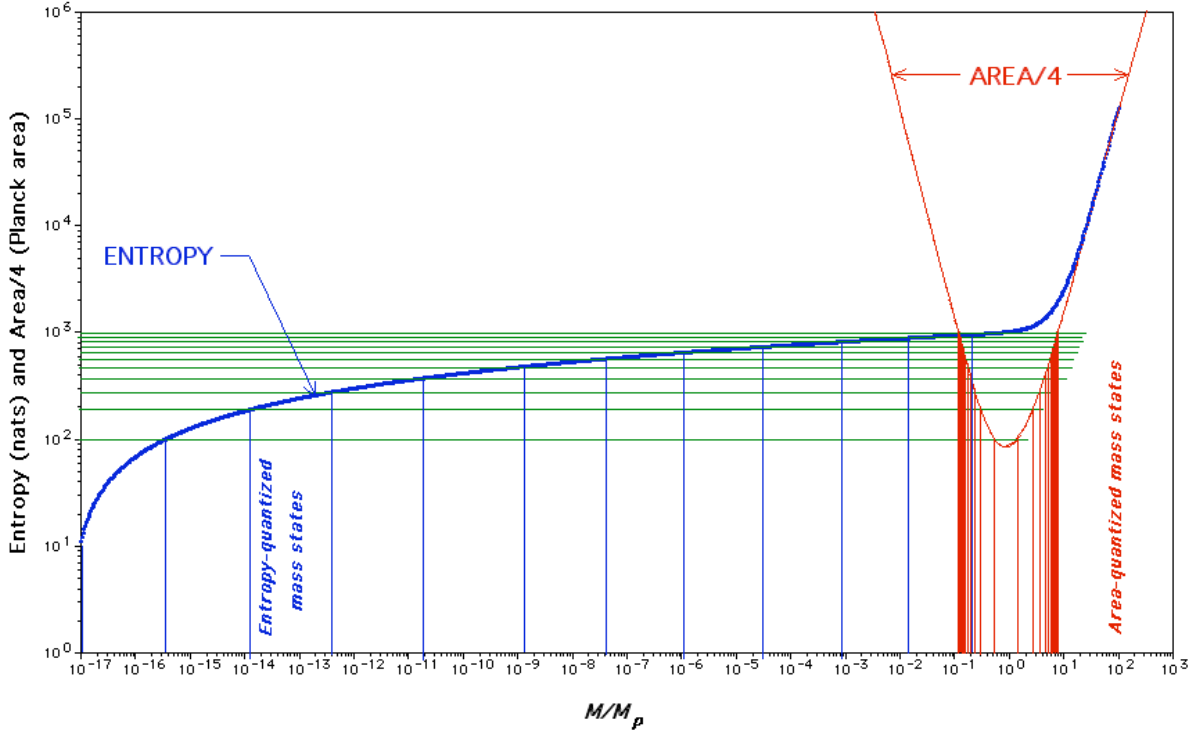

*FIG. 10. A simplified view of the two quantization schemes.*

For entropy-quantization, a casual inspection of Fig. 10 shows that above $M_P$, $S \to A/4$ and the two quantization schemes are indeed equivalent. However, at and below $M_P$ the two schemes are drastically different. The logarithmic form of $S$ requires a minimum mass state $M_L$ in order to have finite and positive entropy, and thus there is a state with quantum number $n = 1$. As $n$ increases, there corresponds a single unique state for each $n$, and the mass spacing of such states is much broader, until the region above $M_P$ where the behavior becomes congruent with the closer-spaced states of area-quantization.

It is implied by Fig. 10 that area-quantization in SSGS makes a very odd, if not untenable prediction for the lowest lying scalar states. The states are already getting very compressed even as the mass is reduced to only 0.1 $M_P$; it is natural to ask if, compressed even further over the next 16 decades of mass reduction, and with reasonably smaller quantum jumps, they start to overlap and make a continuum. The same question will have to be asked for entropy quantization, though the low-mass spectra just shown seem better behaved.

This pictorial scheme for finding the states can be quantified in the following generalization, using the analytic formulae (5.2 and 3.12) for area $A(M)$ and $S(M)$, and their inverse functions, $A^{-1}$ and $S^{-1}$, the latter two evaluated numerically. We use $g_{vp}$ throughout this section (there is no significant difference using $g_{thermo}$ with the standard parameters). For area quantization:

$$A_{\min} + (|n|-1)qA_0 = A\left(M_{|n| \text{ or } -|n|}\right), \quad (5.4)$$

where $A_{\min}$ is not an arbitrary parameter but the minimum value of Eq. (5.2), namely $A_{\min} = 340$

Planck areas (at $m = .87$). The only free parameter, given $A_0$, is $q$, the generalized number of quanta needed to establish a state. The area-quantized states are then formally given by inversion of **5.3**:

$$M_{|n| \text{ or } -|n|} = A^{-1}\{A_{\min} + (|n|-1)qA_o\}, \qquad |n| = 1, 2, 3, \ldots \qquad (5.5)$$

This scheme covers the entire black hole mass spectrum from zero to large astrophysical black holes, with a pair of states for a given $|n|$. The negative-$n$ member lies lower in mass than $A^{-1}\{A_{\min}\} = 0.87$, while the positive-$n$ member lies above $A^{-1}\{A_{\min}\}$. Note that $M_1 = M_{-1}$.
The states of large positive $n$ coincide with those found by previous investigators [35] and are equally difficult to confirm experimentally.

This procedure, and the similar one for entropy-quantization outlined below in Eqs. (5.8) and (5.9), comprise the complete (numerical) solution for area-quantized and entropy-quantized mass states. However, it is useful to explore the quantization analytically in the asymptotic regions of mass (large and small $m$) where $A^{-1}$ and $S^{-1}$ become algebraically simple. We have already done this for high mass area-quantization in Eq. (5.3), and now just substitute the low-mass form for $g$ ($= 1/m^2$) to find the states of large negative $n$ (sub-Planckian):

$$M_n = \sqrt{\frac{4\pi}{(\ln 2)q(|n|-1)}} M_P \qquad (5.6)$$

$$\Delta M \to -\frac{(\ln 2)qM^3}{8\pi M_P^2} \qquad (5.7)$$

…………………………………………………………………………………………..

For entropy quantization we repeat the entire process. First, we define the quantum structure:

$$S_L + (n-1)qS_0 = S(M_n), \qquad (5.8)$$

where $S_L$ is an arbitrary additive constant setting the entropy of the lowest state n=1. We assume it is some multiple $\eta$ of the quantum $S_0$. Each step in $n$ requires $q$ quanta of size $S_0$ to define the next shielding horizon. The lowest permissible mass is $M_1 \approx M_L$, a free parameter of the model, as is $q$. While $S_L$ looks like a parameter, it cancels out in the calculation of the mass states. The entropy-quantized mass states are then formally (and numerically) given by

$$M_n = S^{-1}\{S_o[\eta + (n-1)q]\}, \qquad n = 1, 2, 3\ldots \qquad (5.9)$$

This scheme also covers the entire black hole mass spectrum; the pattern of super-Planckian states of very large $n$ coincides with the states found for quantization of area, and thus with results of other authors. But we are particularly interested in the states which have experimental accessibility, far below $M_P$. For those, the asymptotic form of the entropy is very simple:

$$S = S_L + 8\pi \ln\left(\frac{M}{M_L}\right), \qquad (5.10)$$

and $S^{-1}$ is analytically trivial. Using Eq. (5.9) and inverting Eq. (5.10), one finds:

$$M_n = M_L e^{\left(\frac{(\ln 2)q}{8\pi}\right)(n-1)}, \quad n = 1,2,3\ldots, \tag{5.11}$$

and the gap between two successive states is

$$\Delta M \to (e^{\left(\frac{q(\ln 2)}{8\pi}\right)} - 1)M \tag{5.12}$$

These last two formulae, together with suitable inserted parameter values, will be what we use in section 6 to make experimentally confirmable predictions for LHC.[7]

### B. State widths

While the above stratagems have allowed us to define states, we have no formal theory to tell us what the widths of these states should be. At least a crude estimate is needed, because if the gap $\Delta M$ between two states becomes appreciably smaller than their widths, the spectrum is better described as a continuum. The SSGS does not provide us with a time-dependent wave function, which could be Fourier-inverted to yield the width of states. Instead, all the information of SSGS is lodged in the function $M(t)$, and in the quantized values $M_n$. Estimates of the widths of the states can be made in various ways, and in Appendix C we use two very different methods that felicitously arrive at the same answer.
Our result, however, is algebraically identical to a more sophisticated treatment of Bekenstein and Mukhanov [39], intended for application above $M_P$; we found it to be applicable in both super- and sub-Planckian regions:

$$\frac{1}{\tau} \approx -\frac{dM}{dt}\bigg|_{t_1} \frac{1}{\Delta M_n}, \quad \text{and} \quad \Gamma \approx \frac{\hbar}{c^2 \tau} \;. \tag{5.13}$$

It is not surprising that we get such a definitive answer, while having no detailed knowledge of the partial widths of the original state. Those partial widths would reflect all the ways in which the fundamental Hawking partons could recombine (via their fragmentation functions) to produce more complex final states. In the classical formula for emission power, $dM/dt$, the effects of the multiple modes of *primary radiation* are already incorporated into the generalized Stefan-Boltzman constant $\alpha$, which includes all known particle emission modes as well as spin-dependent gray-body factors [5,6,7]. Thus it is plausible that $M(t)$ represents, via its thermodynamic origins, a statistical average over all such partial widths.

The resulting transition times depend not only on the method of quantization, but in addition on the parameters of the quantization: $M_L$, $q$, $\eta$ and $A_o$ (or $S_o$). The examples quoted below are calculated separately for entropy and area quantization, but typically use a "standard" set of parameters $q = \eta = 16$, $M_L = 100$ GeV, and $A_o$ (or $S_o$) as described. Also note that the mass-dependence of $\alpha$ in a realistic model will modify not only the time scale of $dM/dt$, as noted earlier, but can also modify transition times and widths. Once the mass spectra are actually known, the SSGS model can be reworked to reflect the empirical values of parameters. While we expect the same general picture to hold true, the details will certainly be modified.

---

[7] If q is not a constant, then the model will have no predictive power for mass states unless there is some distinct empirical regularity to the function $q(n)$. For Bohr's atomic physics, the angular-momentum equivalent quantization had $q=1$, $\eta = 0$.

In the super-Planckian region, where we think we know how classical black holes behave, entropy- and area-quantization give lifetime results identical to one another: $\tau \propto M$. By contrast, the classical Hawking *entire* life of a black hole is given by $\mathcal{T} = M^3/3\alpha$. The entire life $\mathcal{T}$ is not a pertinent quantity if one is considering a mass measurement of a particular quantum state, as discussed earlier. If those considerations are included, along with the appropriate $\Delta M_n$, then we find the equivalent Hawking $\tau \propto M$. Ultimately, though, there is no real test of Eq. (5.13), because no experiments are yet capable of detecting these spectral lines from readily available astrophysical black holes.

In the sub-Planckian region, the only basis for comparison is to look for elementary particles that are similar to those of "black hole particles" of equal masses and then compare their lifetimes; this, too, is of somewhat limited utility because no scalars have ever been found, and the results are likely charge and spin-dependent. Eqs. (5.5), (5.9) and (5.13) together show that there is a gigantic difference in the predictions of the two types of quantization for black-hole particle transition times. The widths of these states for a wide range of masses (including the asymptotically low and high ends) are shown by the green curves in Figs. 12 and 13. For instance, at about the TeV mass scale, area-quantization predicts a particle lifetime of order $10^{-57}$s, and one even shorter ($10^{-60}$s) at the GeV mass scale. This clearly has no connection with any experimental physical reality, and of course leads to continuum behavior (to be discussed below).

Entropy-quantization in the sub-Planckian region fares much better, with transition times looking entirely consistent with low-energy experience. At TeV masses, the predicted transition time is of order $10^{-25}$ s, and at a GeV, $10^{-22}$ s. The particle most similar to our uncharged scalars might be the $Z^0$ (neglecting its vector nature), and the prediction of SSGS for the above parameters is for a transition time corresponding to a width of 0.8 GeV. The experimental width is $\Gamma_Z \approx 2.5$ GeV [20], giving some credibility to our transition time calculations. Conversely, if one believes the width estimations are approximately correct, the above set of differences in transition times provides a strong discriminant between entropy- and area-quantization schemes.

Another reassuring exercise is to calculate the width of the lowest neutral scalar $n = 1$ as a function of its mass $M_L$ (a free parameter) and to compare it to a typical minimal supersymmetric model calculation [47] of the same quantity; $M_L$ is, in this example, a Higgs candidate. Fig. 11 shows the result, where the SSGS model matches the MSSM Higgs width (for h and H) very closely for the "standard set" of SSGS parameters and the tanβ=30 case presented by those authors. (For simplicity, we have not included the fast variations of width as H replaces h). There are large possible parametric variations for both models, so this tight agreement is likely fortuitous, but it illustrates again that the SSGS model is giving sensible widths for a scalar boson coupled to mass. We make no implication that the SSGS suggests such a "black hole Higgs" should have other properties related to those of a Higgs particle. In fact, while the Higgs couples only to mass and has a suppressed H—>2γ partial width, the black hole Higgs would presumably couple to mass-energy, like any other object obeying GR, and such a decay would not be suppressed.

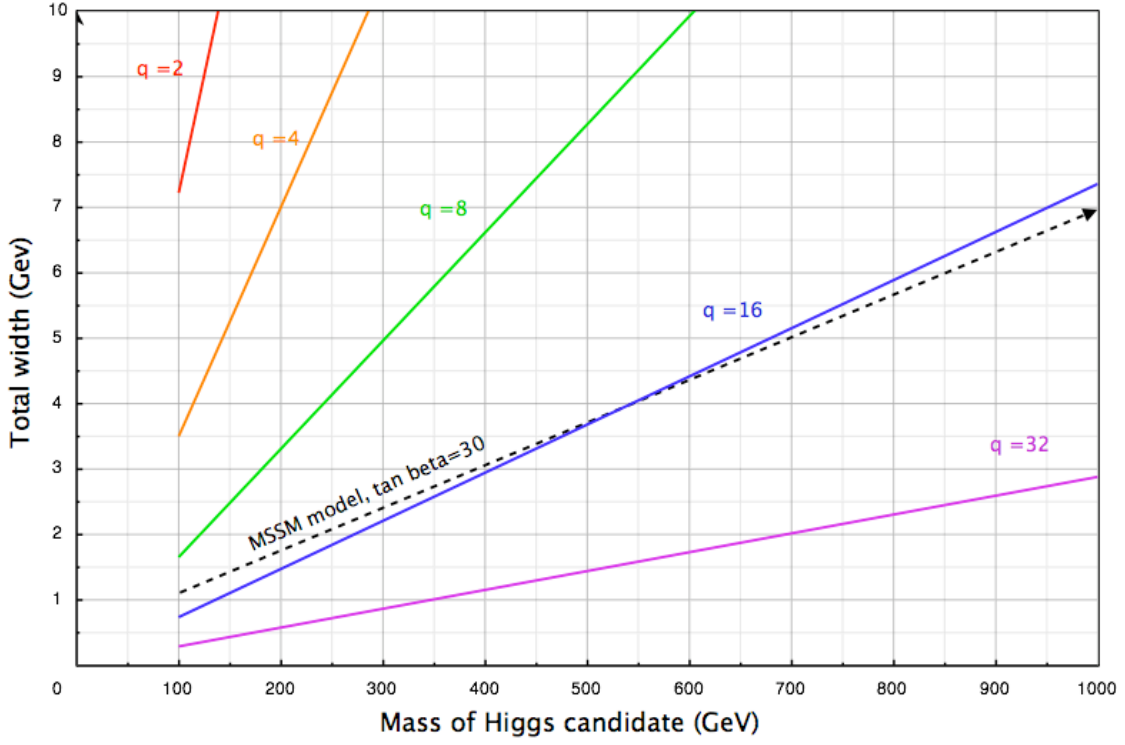

FIG. 11. *Comparison of Higgs width in a MSSM model (tanβ=30) with that of the ground-state scalar of the SSGS model*

Now we are prepared to answer the question as to which quantization scheme, if either, is preferred, given that they are not equivalent for sub-Planckian masses. The answer hinges in part on whether a given scheme yields discrete non-overlapping states, or produces a continuum of unresolvable states. We use the standard set of values for the SSGS parameters as above, which will serve to illustrate the general properties of the solutions we have found.

The results are shown in the rather busy Figs. 12, 13, where we have plotted the dimensionless $\Delta M/M_P$ for the two quantization schemes, plus the previously derived estimate for the widths of states at relative mass $m = M/M_P$. The *n*-value for any quantized mass is also shown for both schemes; one selects integer *n* from the continuous curves.

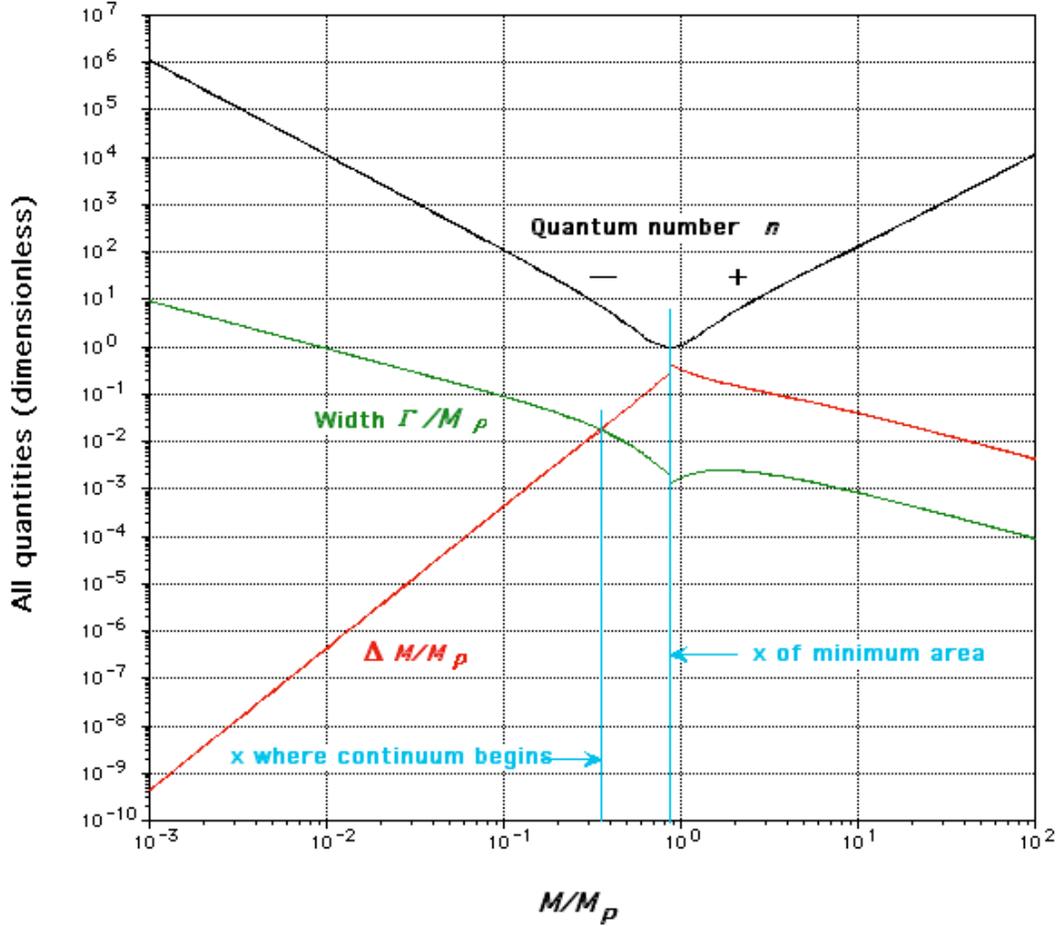

*FIG. 12.  Distinguishability of individual mass quantum states in area-quantization; the discontinuities near m (or x) = 1  are artifacts of the asymmetry of A(m) and the integral nature of n.*

Now it is easy to deduce where states will merge and where they will be distinguishable. The condition for distinguishability is simply that a particular $\Delta M/M_P$ curve must be above the $\Gamma/M_P$ curve. For area-quantization, this is satisfied only if $m = M/M_P > \sim 0.4$. For the choice of $q$ made above, that means only for quantum numbers $-6 < n < +\infty$. Physics at the TeV level would correspond to $n \approx -10^{32}$, where $\Gamma/\Delta M \approx 10^{33} \text{GeV}/4.4 \cdot 10^{-30} \text{GeV} \approx \mathcal{O}(10^{62})$, a rather convincing continuum! It isn't clear to these authors how this continuum almost up to $M_P$ would manifest itself in S-wave amplitudes, but the enormous density of states would suggest a huge cross section; LHC might become a very hot beam dump. This scenario shows little experimental promise at best, and at worst is simply inconsistent with the observed total absence of scalars below $\sim 100$ Gev/c², (there is no minimum threshold). The scheme has no experimentally sensible lifetimes at low energy.  It does say, as expected from our mimicking Bekenstein's work, that in the super-Planckian region, black holes would form semi-stable states, with, e.g., ~1 TeV transitions occurring at $M \sim 7.4 \cdot 10^9$ g, $n \sim 2 \cdot 10^{30}$ and $\Gamma \sim 0.02$ TeV.  Still, since this scenario appears unredeemable in the sub-Planckian region, we pursue it no further in this paper.

In Fig. 13, we find that a parallel examination reveals a quite different prospect for the scenario where we quantize entropy.  It gives the same plausible result just discussed for the super-Planckian region, where it is indistinguishable from the classic theory (except for the labeling of the index *n*). But in the sub-Planckian region, because of the logarithmic behavior of the entropy there, the quantum states are stretched out. Fig. 13 reflects this, in that it shows a radically different behavior for *n(m)*, and because $\Delta M/M_P$ (entropy) is *always* well above the $\Gamma/M_P$ curve. The quantized mass states do not form a continuum anywhere (at least not from intrinsic width effects), below or above the Planck mass. In fact, this analysis predicts that the widths of the neutral scalar states, for low-lying entropy-quantized masses, are a constant 0.74% of the observed mass; the gaps between states are ~56% of the observed mass. Thus $\Gamma$ is about 1.3% of the gap size $\Delta M$ and there is no continuum region. This holds all the way down to the lowest state at $M_L$, required for self-consistency of the model.

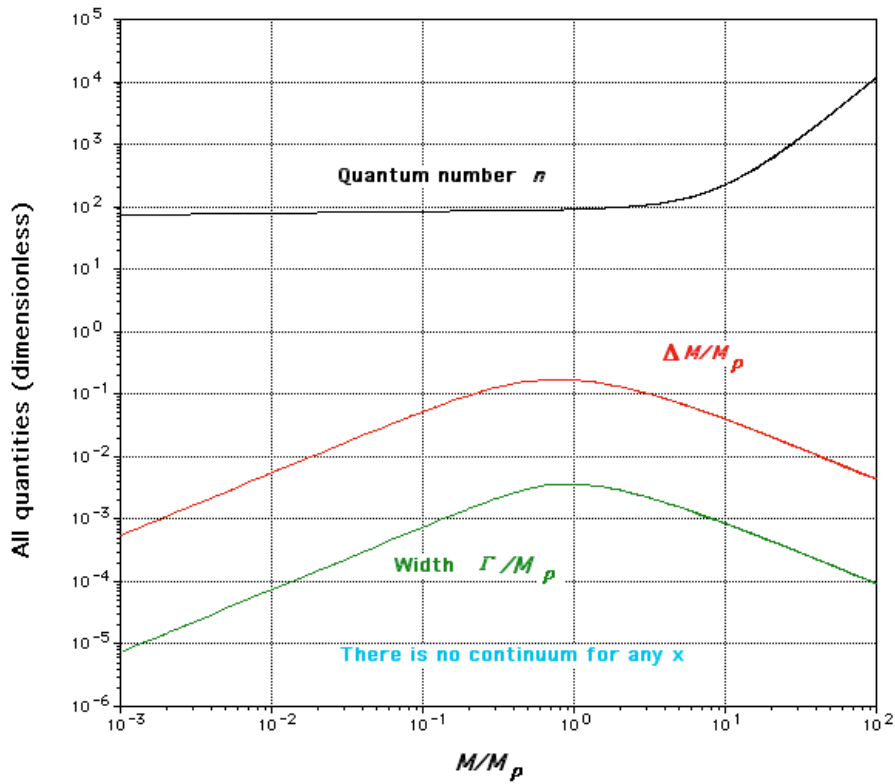

*FIG. 13. Distinguishability of mass quantum states in entropy-quantization.*

## VI. CONFRONTATION WITH EXPERIMENTS AND OTHER THEORIES

Given this encouraging picture for the entropy-quantized scheme, we offer some predictions for physics at LHC or higher energies. Fig. 14 shows individual low-lying states for the parameters given, drawn with their calculated $\Gamma$'s when such $\Gamma$'s are greater than the minimum line width available for the drawing. For completeness, the behavior of the entropy-quantized states near the Planck mass is also shown in Fig. 15, where only a few characteristic *n* are shown to simplify the diagram. Note that there are "only" 91 states from 100 GeV up to the Planck mass for this particular parameterization. While that may seem like a plentitude, high-energy physics has explored only about $10^{-17}$ of the possible spectrum up to $M_P$. The spectral density of particles already found is enormous compared to what this example predicts for higher energy regions. To future experimentalists, those regions may indeed seem like a desert.

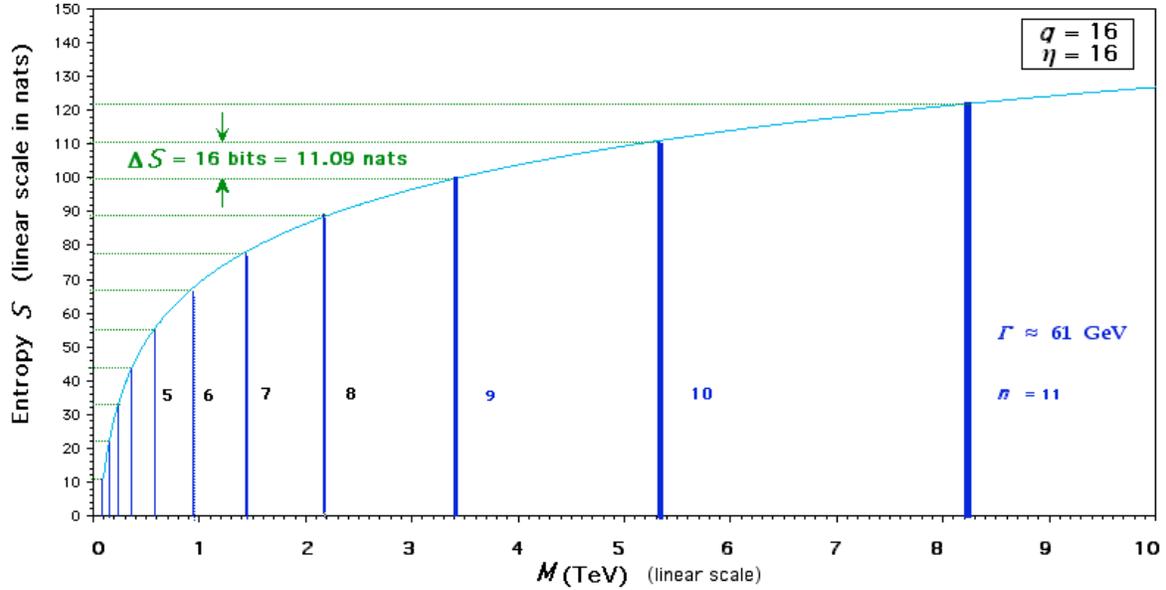

FIG. 14.  *Entropy quantization for the case $\eta = 16$, $q = 16$.  Mass states are shown at absolute masses and (where visible) with their predicted line widths.*

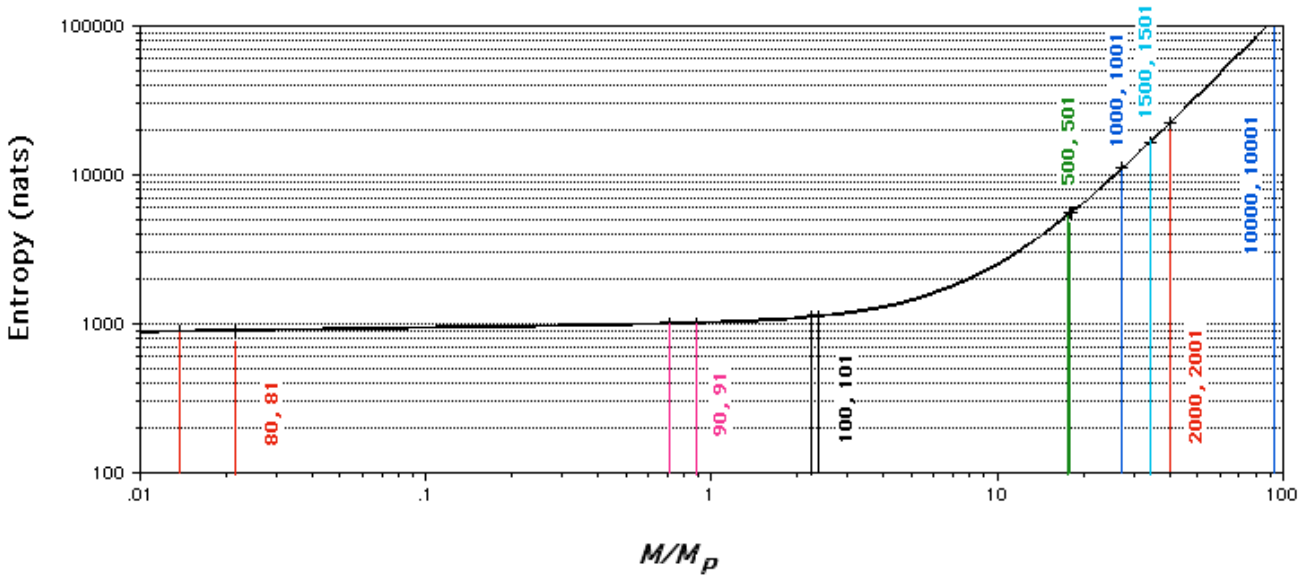

FIG. 15. *Mass state separation in the vicinity of $M/M_P = 1$ for entropy-quantization similar to Fig. 14; numbers by lines designate pairs with adjacent index n.  Widths of states are negligible relative to gaps between n and n+1, even as states seem to merge on the diagram.*

For more detail, Table I below shows the array of lowest-lying accessible states of uncharged scalars to be expected at CERN, if $M_L$ has been correctly guessed at 100 GeV and the required number of fundamental quanta per state is 16.   The usual but arbitrary selection $\eta = 16$ is used, indicating that the ground state also requires 16 bits of information.  We stress that this is an example of the *pattern* to be expected, rather than some particular masses.  ***The real prediction of the SSGS is the exponentially increasing pattern of masses stemming from the logarithmic dependence of entropy.***

| $n$ (q =16) ($\eta$ =16) | $n'$ (q =1) ($\eta$ = 16) | $M_n$ (GeV) | $M_{n+1} - M_n$ (GeV) | $\Gamma_{SSGS}$ (GeV) | Entropy (includes $S_L$) (bits) | Entropy (includes $S_L$) (nats) |
|---|---|---|---|---|---|---|
| 1 | 1 | 100 | 55 | 0.74 | 16 | 11.1 |
| 2 | 17 | 155 | 87 | 1.14 | 32 | 22.2 |
| 3 | 33 | 242 | 131 | 1.78 | 48 | 33.3 |
| 4 | 49 | 376 | 208 | 2.77 | 64 | 44.4 |
| 5 | 65 | 584 | 324 | 4.30 | 80 | 55.5 |
| 6 | 81 | 908 | 504 | 6.68 | 96 | 66.5 |
| 7 | 97 | 1412 | 783 | 10.4 | 112 | 77.6 |
| 8 | 113 | 2195 | 1218 | 16.2 | 128 | 88.7 |
| 9 | 129 | 3413 | 1893 | 25.2 | 144 | 99.8 |
| 10 | 145 | 5306 | 2943 | 39.1 | 160 | 110.9 |
| 11 | 161 | 8249 | 4576 | 60.7 | 176 | 122.0 |
| 12 | 177 | 12825 | 7114 | 94.2 | 192 | 133.1 |

*Table I. The lowest-lying states of neutral scalar black holes, in an example of the entropy-quantization scheme of the SSGS model. If the states are fine-grained with only one bit per level, n' shows the corresponding quantum number.*

For any $M_L$ and any $q$, the generalized Eqs. (5.11), (5.12) tell us, given the first two states $M_1 \equiv M_L$ and $M_2$ from experiment, where the rest of the expanding tower of particles would lie. The spacings will look linear in the mass only for a few lowest-lying states, because both the mass levels and their spacings depend *exponentially* on $n$. The SSGS model thus intrinsically admits a mechanism suggesting that families of elementary particles can show gigantic differences in mass. In SSGS, the quantity $q$ is the number of fundamental bits of information needed to specify the next higher shielding horizon. If $q$=1 and only one bit is needed (which seems unlikely), then there are ~1500 uncharged scalar states between 100 GeV and the Planck mass; if $q$ =100, there are only ~15. In this latter case, if $M_L$ = 100 Gev, $M_2$ would be at ~1.6 TeV and the pattern of higher states would be less accessible at LHC in 2008. The main example illustrated in the table, $q$ = 16, is more felicitous for significant testing at LHC.

In addition to the prediction of quantized neutral scalar states, once LHC can detect above the threshold $M_L$, we predict a generic high energy behavior remarkably different than the speculation about "black hole factories at LHC". A variant of Fig. 16 is often used by various lecturers [48] to illustrate a classic prejudice: why black holes can't turn into elementary particles. The reason: no black hole can become smaller than its Compton wavelength, and the classical black hole trajectory does. For SSGS, Fig. 16 is a compact way to dismiss this prejudice and to show the substantial differences SSGS has with the "black hole factory" scheme [2] of string theory. The trajectory of an evaporating black hole in SSGS undergoes a smooth but very rapid transition near $M_P$,—from following the GR lower limit of size—to following the scale dictated by the Uncertainty Principle for meaningful, measurable sizes of objects in QM: the Compton wavelength. No variance with our concept of particles exists for these SSGS black holes. On Fig. 16 we have also plotted, for purposes of completeness, an alternate dynamic trajectory for black hole evaporation derived earlier by Adler, et. al. [49], using their formulation of a Generalized Uncertainty Principle(GUP) [50]. That solution had the curious property of suddenly terminating at the Planck mass, even though the hole was extremely hot and radiating profusely. We took the liberty of substituting our variable $G$ into their analytic solution, and found that then their solution does not "freeze" at $M_P$, but becomes similar to our SSGS solution to within < 2% over the entire mass range.

We have been discussing dynamic trajectories following the evaporation of a black hole downwards in mass. At an accelerator, however, we usually view the process using data gathered with increasing c.m. energy $E$ of the interaction involved, looking not only for new mass states but for the general behavior of cross sections, multiplicities and geometrical distributions, believing that such data can characterize internal properties (temperatures, entropies) of the black holes if they are produced [3,4,51]. The insets of Fig. 16 show that these properties have energy dependences quite different for SSGS (in the energy regime of LHC) compared to the classical black hole region, and compared to the string theory/brane extrapolations that predict black hole factories [51]. In particular, except for the resonant behavior of new particles, the LHC cross sections σ should continue to decline like $1/E^2$ instead of increasing like $E^{1/3}$, and the effective black hole temperature (if deducible) should rise linearly with E instead of falling like $E^{-1/6}$. Entropy in SSGS will be logarithmic instead of rising like $E^{7/6}$. (The string theory values depend on the number $D$ of *spatial* dimensions in excess of the conventional three; we have used the case $D=6$, implying a string theory with ten spacetime dimensions). With adequate experimental techniques, these differences should be manifest. The only caveat we can see is that LHC may well be operating just below the $E$ where the string theory model "leaves" the Compton trajectory—the position of the lower "question mark" in Fig. 16. There, the string model demands a full QG treatment to describe the objects produced [51], unlike the calculable properties for a black hole well above this threshold and on the dashed trajectory in Fig. 16. In SSGS, there are no such caveats; everything is expected to change smoothly.

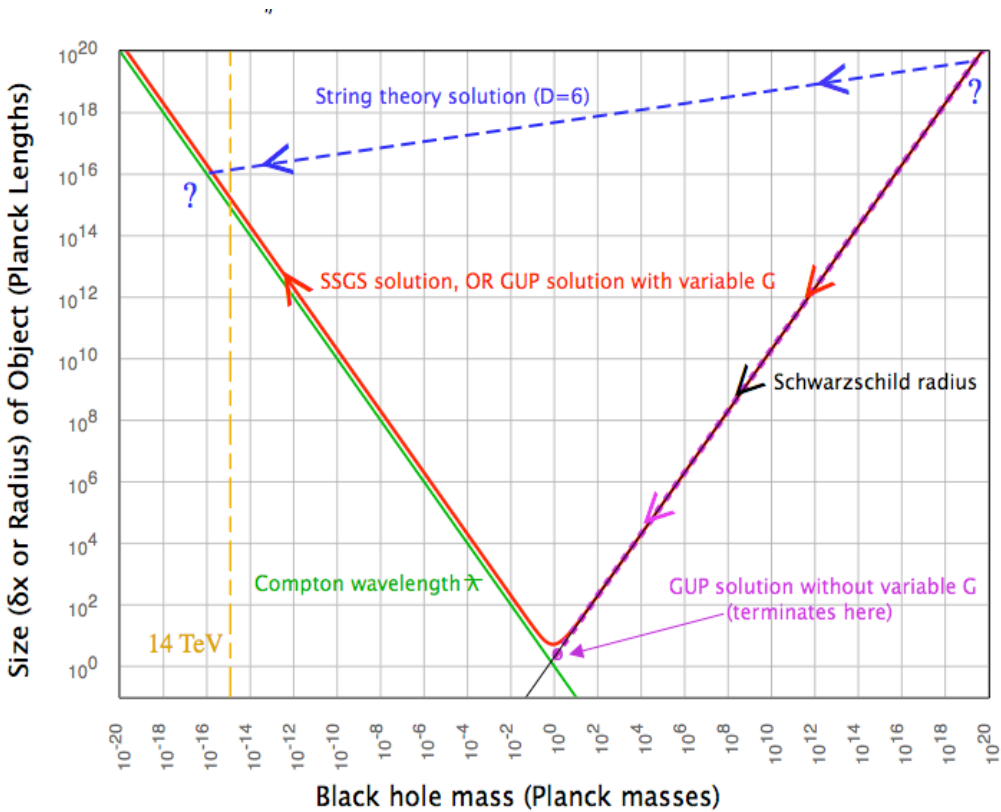

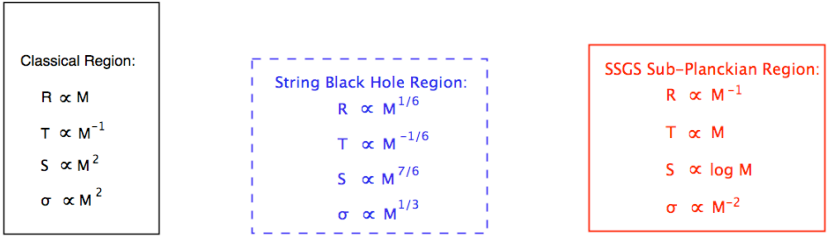

FIG. 16. Comparison of behavior of models of black hole production at LHC (see text)

## VII. DISCUSSION AND CONCLUSION

At first glance the scenario derived in SSGS seems bizarre, but it is not: this is exactly what would be expected if an evaporating black hole leaves a remnant consistent with quantum mechanics. One might posit that the black hole smoothly turns into something approximating a large and unstable elementary particle, which then continues to evaporate (decay) into familiar stationary states. Many have speculatively done so: some models suffered from unrealistically small horizon sizes, and some have found ingenious ways to avoid them [52]. No model has provided reliable predictions of particle spectra. One early bold speculation put the matter quite simply:

> At the Planck scale it may well be impossible to disentangle black holes from elementary particles. There simply is no fundamental difference. If black holes show any resemblance with ordinary particles it should be possible to describe them as pure states, ... The spectrum of black holes states is discrete. Baryon number conservation, like all other additive quantum numbers to which no local gauge field is coupled, must be violated [8].

With repeated attempts by various authors, this speculation became more quantified and robust:

> We suggest that the behavior of these extreme dilaton black holes….can reasonably be interpreted as the holes doing their best to behave like normal elementary particles [9].

What the SSGS seems to suggest (using a much more naive theoretical structure than the above papers) is that *all* particles may be varying forms of stabilized black holes, and that such objects have appreciable *intrinsic* entropy in addition to any kinematical thermal entropy they might possess. This would put a whole new light on the process of evaporation of large black holes, which might then appear no different in principle from the correlated decays of elementary particles; there would be no need for infinite reservoirs of information. The entropy (and information) could reside just as suggested in [26]—in the deviations of the black hole "atmosphere" from complete thermality—or, more to the point of this paper, in the structure of the shielding horizons. The large-hole evaporation calculations would have to include a possible flow of correlated information via the intrinsic entropy of the "daughter holes". Without further speculations, we have already seen a big hint that this can happen, namely, in the calculations of Appendix B that showed it is possible that the new intrinsic entropy alone can satisfy the second law of thermodynamics, for both large and small hole evaporation. That would depend on a larger conjecture being true: that all particles are black holes, not just the uncharged scalar modes.

As of yet, the SSGS has no mechanism to explain the details of the large-$G$ horizon membrane or of a leaky horizon. It suggests, however, that a consideration of unconventionally strong $G$ behavior exceeding that of GR might automatically provide a sensible environment and "gentle landing" for the evolution of an evaporating black hole into any scalar fundamental particle, consistent with QM everywhere and GR on the exterior of the hole. From there, it is only a short flight of fancy to include all elementary particles as stabilized black holes.

The "bottom line" of this discussion of SSGS is revealed by inspecting the value of $M_{P(vp)} = \sqrt{\hbar c / G}$ in the low mass asymptotic region (where real particles are). One can again apply the $G$-scaling rule and finds:

$$M_{P(vp)} \rightarrow M \qquad (!) \tag{7.1}$$

If black holes become particles by the mechanism herein proposed, then the Planck scale pertinent to them is their *own mass* at which they quantize and stabilize. In other words, the mechanisms of QG are already active at *every level* in which we talk about elementary particles.

This discussion concludes with the proffered answers to the objections raised in section 3 and elsewhere in the paper. These answers are purely speculative, and we make no claim that they follow directly from the SSGS phenomenological model.

1) Gravity is seen everywhere in the Universe because all particles, conjectured to be bound black hole states of intense gravity, leak the weak asymptotic gravity (GR) into their environs; unstable small black holes and the explosions of the sort sought by experimenters are not necessary for the scenario.

2) Many suggest that gravity is an un-shieldable force because there is no matter with repulsive gravity. First note that traditional weak gravity is not shielded in this model, and only a finite strength of strong gravity need be shielded. It is known that about 70% of the universe is made up of a "dark energy" which is very dilute and seems equivalent to a cosmological constant—effectively a repulsive gravity. Is it really certain that the ~4% ordinary matter and the ~25% dark matter could not be shrouded by locally intense concentrations of this unknown quantity at the elementary particle level? With coarse-graining this might not be noticeable in cosmology, though in fine-grained experiments this mechanism might be falsifiable.

3) As for string theory successes, this paper certainly does not suggest they are wrong or fortuitous; it suggests only that the key elements are the number of degrees of freedom of that theory and the equations, not the interpretation of those degrees of freedom as arising from unseen dimensions, nor the equations as describing real strings. As mentioned, the real shielding mechanism of the strong interactions of QCD arose from the color degrees of freedom and the non-Abelian structure of the theory. It is possible that the freedom provided by the extra dimensions of string theory map one-to-one onto new aspects of a quantum theory of strong gravity with shielding horizons, and would then provide a unique solution for string theory instead of the proposed "Landscape" of solutions.

4) Finally, as for the future LHC experiments, they may very well sample a region of T where semi-stable black holes of M ≈ T will be produced, but, in the scenario presented in this paper, that is what high-energy physics has been doing all along at other energy scales.

In SSGS, particles/holes decay by passing entropy and information on to the daughter particles/holes, which suggests that strong correlations will exist in the decay products. From the SSGS viewpoint, this has always been true in high-energy physics, where every particle repeatedly sampled the appropriate variable Planck "endpoint" for its particular mass. The scenario offered by this paper, unlike those of [2], suggests that the end to conventional high-energy physics is comfortably far away.

## Acknowledgements

Jonathan Hanna was very helpful in performing some preliminary numerical integrations as we searched for viable models [53]. We thank Fred Kuttner and Bruce Rosenblum for their critical readings of the archival reports, Melanie Mayer for her attention to our prose and Benjamin Lubin for his technical advice in the preparation of manuscripts. We thank Ronald Adler for insights to works on GUP. Tom Banks has contributed his considerable insights on matters of entropy and holography; he is not responsible for our unorthodox viewpoints.

## APPENDIX A: ASYMPTOTIC FORMS OF SSGS

The following relationships all follow from definitions, derivations and equations in the main text.

**Asymptotic forms, super-Planckian region:**

$$t - t_0 = -\frac{M_P^3}{3\alpha}\left(m^3 - m_0^3\right) \quad \text{(The Hawking formula)}$$

$$\frac{G}{G_N} \equiv g = 1$$

$$A = 16\pi m^2, \text{ in Planck units } \frac{\hbar G_N}{c^3}$$

$$S = 4\pi m^2$$

*Interrelationships (to show duality properties with sub-Planckian region):*

Define $r = R/R_P$, where $R_P$ is the horizon radius of a classical Hawking black hole at $M_P$, and $T_{NORM} = T/T_{PH}$, where $T_{PH}$ is the corresponding Hawking temperature at $M_P$. Then for $m \gg 1$,

$$m = r = \frac{1}{T_{NORM}}$$

**Asymptotic forms, sub-Planckian region:**

$$t - t_0 = \frac{M_P^3}{\alpha} \frac{1}{m}$$

$$\frac{G}{G_N} \equiv g = \left(\frac{1}{m}\right)^2$$

$$A = 16\pi \left(\frac{1}{m}\right)^2, \text{ in Planck units } \frac{\hbar G_N}{c^3}$$

$$S = S_L + 8\pi \ln\left(\frac{m}{m_L}\right)$$

*Interrelationships (to show duality properties with super-Planckian region):* For m << 1,

$$\sqrt{g} = \frac{1}{m} = r = \frac{1}{T_{\text{NORM}}}$$

# APPENDIX B: A CONJECTURED SECOND LAW OF THERMODYNAMICS FOR INTRINSIC ENTROPY

We wish to calculate the net gain or loss of entropy in a black hole-thermal environment when the black hole evaporates, counting only the intrinsic entropies possessed by the original hole and the emitted particles. If we were to do this classically, we would find a marked *decrease* in the total intrinsic entropy, as exemplified by the following simple case: a black hole mass of $M_o$ completely evaporates into $N$ identical black holes at rest, each of mass $M_f$ with $N = M_o / M_f$. Tracing the intrinsic entropy $S_o = 4\pi G_N M_o^2 / \hbar c$, $S_f = N(4\pi G_N M_o^2 / N^2)/\hbar c$, we see that $S_f / S_o = 1/N$. So for any $N > 1$, the intrinsic entropy decreases, and for the usual case where $M_f << M_o$, it decreases enormously.

In the real world, large black holes are not expected to bifurcate or polyfurcate into moderate sized holes, but instead into particles *elementary at the scale of the large black hole temperature.* Fig. 1 showed what this means quantitatively: e.g., a black hole of temperature 1 TeV doesn't Hawking-radiate into pions, but into quarks and gluons. A solar-size black hole emits only photons or other zero-mass particles. Fig. 17 quantifies this further, showing an interesting duality symmetry we noticed long ago [36,56] between the mass of holes and the heaviest particle into which they can decay. One sees immediately that the simple example above has no realization unless $N$ is gigantic. Moreover, even though GR and the variable-pole model give identical results for the intrinsic entropy in the large mass region, the *flow* of entropy will be entirely different for the two schemes, because the larger masses reference the sub-Planckian region where the treatment of GR/Hawking objects and objects from the SSGS model has to be completely different.

Thus, in the case of Hawking evaporation of a large hole into classical elementary particles, the only entropy flow is from the intrinsic large-hole entropy into the non-intrinsic entropy of fundamental particles (reflecting the lack of kinematical information about their micro-canonical ensembles). In our new SSGS viewpoint, the final state particles have intrinsic entropy too, in addition to kinematical information content. This intrinsic entropy could in principle be entropy of entanglement, i.e., it could be combined with the entropy of the evaporating black hole and other particles to keep the combined state a pure one. (This assumes the original black hole was in a pure state). For this view to be a real possibility, the intrinsic entropy in the final state should *increase* throughout most of the evaporative process [57,58], which is just the opposite of what we found in the simple classical case. So, what really happens?

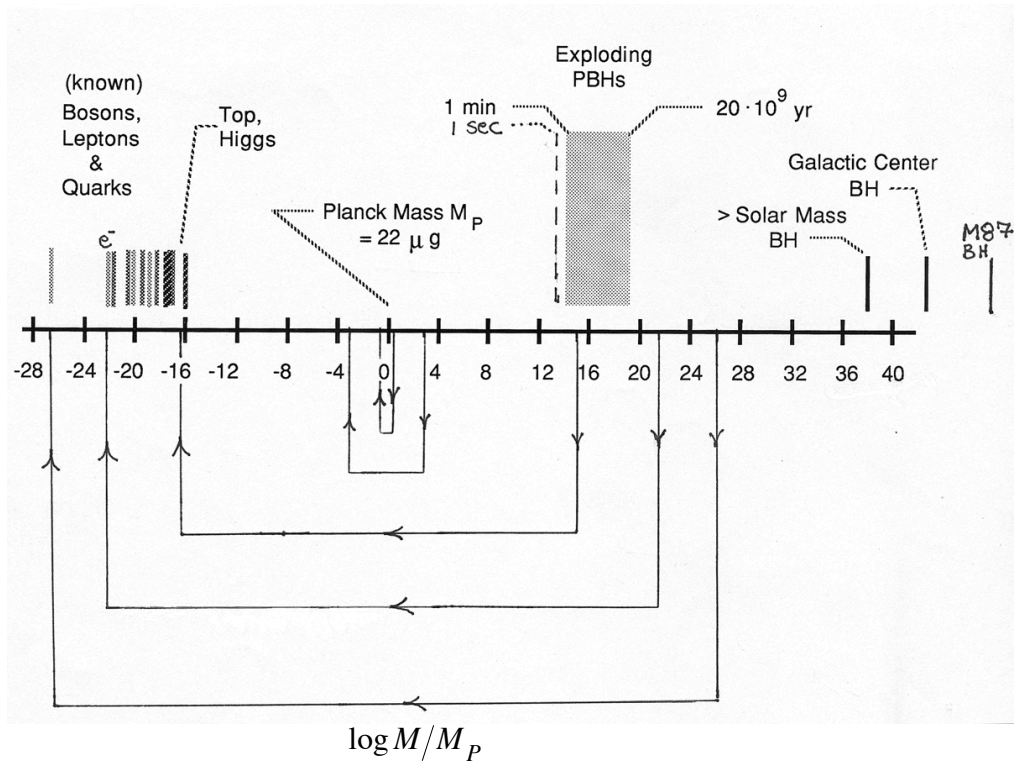

*FIG. 17. The landscape of masses M: the arrows indicate the most massive particle (at left) that can be emitted by a black hole (at right).*

A proper calculation of the evaporation process, whether it starts at super-Planckian or sub- Planckian masses, is beyond the scope of this paper. It would involve a large Monte Carlo program as described in [59], modified for the variable-pole assumptions—convolving all possible particle distributions (production, fragmentation, decay) with the Hawking radiation equation. We can, however, do an approximate calculation more sophisticated than the trivial, but revealing one, with which this Appendix began.

A key approximation, justified by [7,59], is that one can approximate the Hawking thermal spectrum by a δ-function at the peak energy of emission, $\sim 5k_B T = 5\kappa k_B/MG_N$, which then becomes the typical energy per emitted particle, or, more crudely, the total mass of each emitted "black hole fundamental particle". Then, instead of a one-step process, we take the more precise view of the evaporation proceeding differentially. We know $dM/dt$ and $S(M)$, so for each step of $\Delta M$ it is easy to calculate how much entropy is lost from the "mother" black hole. After the loss of $\Delta M$, intrinsic entropy from the mother hole decreases, but then we have a known average number/time of "daughter holes" $(dM/dt)/(5k_B T)$, each with a new intrinsic entropy given by $S(M_{daughter})$. Knowing $M(t)$, we then can deduce how much new entropy has appeared in the daughters during $\Delta M$. Note that we have neglected further decays of the daughter particles because they are already fundamental and are stable, or at worst decay into a few stable particles that have negligible intrinsic entropy. The only subtlety in this calculation is if the mother black hole is already sub-Planckian. If so, its temperature is then set by the duality conditions of Appendix A, and it has the same temperature of a super-Planckian hole of $1/m$. Thus the sub-Planckian mother can decay into precisely the same set of daughter holes as could its super-Planckian dual, a somewhat surprising but sensible result.

In Fig. 18 we display the pertinent result of integrating the sum of the differential changes in intrinsic entropy as evaporation proceeds:

$$S_{total} - S_{initial} = \int_{M_0}^{M} \left( \Delta S_{gained\ in\ final\ state} - \left| \Delta S_{lost\ from\ M} \right| \right)$$

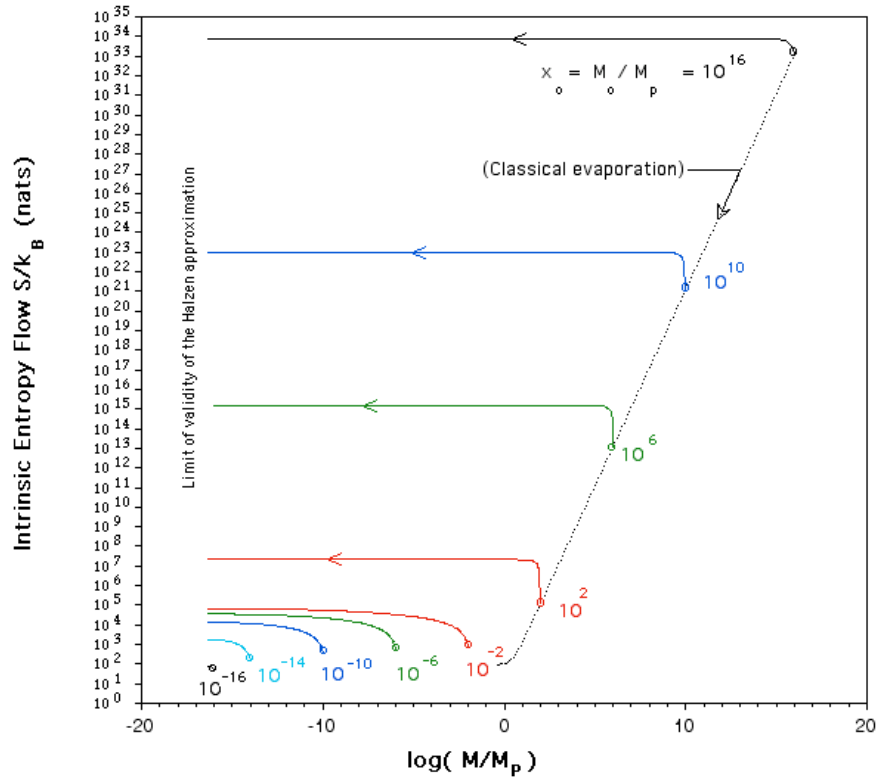

FIG. 18. *The change in total intrinsic entropy as a SSGS hole evaporates: given parameter is initial mass; classical solution shown by dotted line.*

Shown is $S_{total}$, using a variety of initial masses $M_0$. ($S_L$ is taken to be zero, but even with $S_L = 100$, the general nature of the curves in Fig.18 does not change—the saturation values for the total entropy just rise somewhat higher). One sees that the evaporation starts at the right of each curve, with the initial intrinsic entropy marked by a small circle. In every case for mother holes of super-Planckian mass, the total intrinsic entropy rises sharply as the evaporation proceeds along the direction of the arrows. The entropy saturates quickly for all cases except for mother holes of sub-Planckian masses, where the rise and saturation occur more slowly. The integration is terminated (at the left side) when the peak energy of Hawking radiation becomes smaller than $M_L$ and the Halzen approximation breaks down. By this point the remaining mass and entropy of the mother black hole are very small, and the evaporation is best described by ordinary particle decay.

By contrast, if we still use the same approximation, but say that all black holes— including sub-Planckian ones—have the classic intrinsic entropy accorded to large black holes, we get the dotted trajectory for total entropy shown in Fig. 18. The Second Law, taken for intrinsic entropy alone, is violated quite severely. It is the logarithmic behavior of particle intrinsic entropy that is "saving" this specialized Second Law. Also note that just as the mother hole is rapidly shedding its entropy, proportional to $M^2$, the daughter particles go off with even more than was lost. We conjecture that all this behavior will still be true if the model were generalized to include vector and spinor black holes, eliminating any case that would have information loss in the evaporative process. It is perhaps unwise to pursue the model in so much detail. But overall, the conclusion is simply that there seems to be a great deal of previously uncounted intrinsic entropy in this model.

## APPENDIX C: DECAY RATES OF QUANTIZED STATES

**Method 1 (statistical):** We interpret $M(t)$ to be a statistical average over a large variety of quantized modes of decay. Using a simple two-state system of masses $M_1$ and $M_2$, what we can physically observe in a single trial (ST) is:

$$M_{ST}(t) = M_2 + (M_1 - M_2)H(t_d - t), \qquad (C.1)$$

where $H(y)$ is the Heaviside function ($H(y) = 1$ if $y > 0$, and $H(y) = 0$ if $y < 0$); $t_d$ is the time at which $M_1$ decays into $M_2$. Averaging $M_{ST}$ over the decay probability distribution for $M_1$ (with lifetime $\tau_1$),

$$P(t_d) = \frac{1}{\tau_1} \exp\left(-\frac{t_d}{\tau_1}\right), \qquad (C.2)$$

we find the statistically-averaged observable:

$$M(t) = M_2 + (M_1 - M_2)e^{-t/\tau_1}, \qquad (C.3)$$

and by differentiating and manipulating the terms we arrive at the result:

$$\frac{1}{\tau_1} = -\frac{dM}{dt}\bigg|_{t_1} \frac{1}{M(t_1) - M_2} = -\frac{dM}{dt}\bigg|_{M_1} \frac{1}{M_1 - M_2} \; . \qquad (C.4)$$

**Method 2 (heuristic):** The previous method does not include multiple states with varying lifetimes. If we instead interpret $\tau$ as a mean transition time from a given state to the next lowest state, and assume that $dM/dt$ reflects the classical limit as the mean value of radiated power (black hole energy loss per unit time), then $dM/dt = -(M_n - M_{n'})/\tau$ for two closely-lying states, resulting in the same form as method 1, for two closely-adjacent states. This seems correct in another sense: What limits the length of an interval δ*t*, in which we sample a mass described by *M(t)*? We can make the interval arbitrarily short and artificially induce a large width, but is there any *upper* bound to the interval? We know that the interval δ*t* samples a mass range δ*M* = -(*dM/dt*) δ*t*. For the measurement of the mass to be pertinent to the mass $M_n$, δ*M* must be smaller than the gap Δ*M* between quantized states. This line of argument gives an estimator for the inverse transition time:

$$\frac{1}{\tau} \equiv \frac{1}{\delta t} \approx -\frac{dM}{dt}\bigg|_{t_1} \frac{1}{\Delta M_n}, \quad \text{and} \quad \Gamma \approx \frac{\hbar}{c^2 \tau} \quad . \tag{C.5}$$

This constitutes a lower limit for *Γ*, which in principle could be attained, suggesting it is a good estimate available for the natural width.